\documentclass[multphys,vecphys]{promult}

\usepackage{makeidx}     
\usepackage{graphicx}    
\usepackage{multicol}    
\usepackage{wrapfig}
\usepackage{epsfig}
\usepackage{amsfonts,longtable}
\usepackage {amsmath}
\usepackage{graphicx}

\makeindex             
\begin{document}

\title*{Transient detections and other real-time data processing from MASTER-VWF wide-field
cameras.}

\author{Evgeny Gorbovskoy \inst{1}, Kirill Ivanov \inst{4}, Vladimir Lipunov
\inst{1}, Victor Kornilov \inst{1},Alexander Belinski \inst{1} , Nikolaj
Shatskij \inst{1}, Dmitry Kuvshinov \inst{1}, Nataly Tyurina \inst{1}, Pavel 
Balanutsa \inst{1}, Vadim Chazov \inst{1}, Artem Kuznetsov \inst{1}, Petr
Kortunov \inst{1}, Andrey Tlatov \inst{2}, Alexander Parkhomenko \inst{2}, Vadim
Krushinsky \inst{3}, Ivan Zalozhnyh \inst{3}, Alexander Popov \inst{3}, Taisia
Kopytova \inst{3},  Sergey Yazev \inst{4}, Alexander Krylov \inst{1}
}
\institute{
1 Moscow State University, Sternberg Astronomical Institute,
119991, 13, Univeristetskij pr-t, Moscow, Russia\\
2 Kislovodsk Solar Station, 357700 p.o. Box 145, 100, Gagarina st., Russia\\
3 Ural State University, 620083, 51, Lenina pr-t, Ekaterinburg, Russia\\
4 Irkutsk State University, 664003, 1, Karl Marks st., Irkutsk, Russia }

\maketitle
Construction of robotic observatories has developed into an important and thriving fields 
of astronomy. Their large field of view combined with the capability to be pointed at 
any direction make robotic astronomical systems indispensable for tasks involving searches for 
transients like grb, supernovae explosions, novae etc., where both the time and direction
of the search are impossible to predict. This paper describes prompt GRB
observations made with MASTER-VWF wide-field cameras and the methods of image analysis and
classification of transients used for real-time data processing. 
During seven months of operation six synchronous observations of gamma-ray bursts have been made  
using MASTER VWF facilities deployed in Kislovodsk and Irkutsk.  In all cases high upper limits have been obtained 
(see Table~\ref{tab_grbwf} and Fig. ~\ref {allgrb}).

\section{Introduction}
\label{sec:1}

\begin{table}[t]

\begin{longtable}{ p{4cm} p{2.3cm} p{2.5cm} p{3.5cm} }

\caption {Comparative characteristics of super-wide field systems} \endhead
\hline \hline
Project & FOV & Exposure & Limiting magnitude \\
\hline
 WIDGET\cite{widget} &  $44^{o}$ x $44^{o}$ x 3 & 5 sec & $10^m$  \\
 RAPTOR & $40^{o}$ x $40^{o}$ & 60 sec & $12^{m}$ \\
 Pi of the Sky\cite{pi_of} & $22^{o}$ x $22^o$ & 10 sec &  $11.5^{m}$ \\
 Yatsugatake Cameras & 85x70 & 8 & $5^{m}$ \\
 FAVOR\cite{favor} & 16x24 & 0.13 & $11.5^{m}$ \\
 TORTORA\cite{tortora} & 24x32 & 0.13 & $10.5^{m}$ \\
 \hline
 MASTER-VWF(1)\cite{master} &  $18^{o}$x$25^{o}$ or $28^{o}$ x $ 45^{o}$ & 0.3 -
10 sec & $12^{m}$(5-sec) $9.5^m$(0.3 sec)  \\
 MASTER-VWF(4)\cite{lipunovmaster} & 4x$28^{o}$x$42^{o}$ & 0.3 - 10 sec &
$11.5^{m}$(5 sec) $9.5^m$(0.3 sec)  \\
\hline\hline
\end{longtable}
\label {tab_wf}
\end{table}

\subsection{ Very wide field telescopes. }

Currently, several dozen robotic telescopes are operated worldwide, each of them usually meant
for certain dedicated tasks. A special case of these instruments are super-wide field telescopes with  fields of 
view ranging from hundreds to thousands square degrees. It is impossible to make a very wide field system with a
large aperture using now available CCDs and optical systems, and  that is why the apertures of very wide field cameras
are limited to 5-15 cm. All very wide field cameras are designed to record various optical transients.  
Most  of the optical transients are of circumterraneous origin --- meteors and satellites, however, very wide-field
cameras
may also provide valuable data on the prompt emission of cosmological grbs. A striking example
is the incredibly successful observation of GRB080319B that was performed within the framework of the
Russian-Italian experiment TORTORA \cite{racusin}  and Polish experiment "Pi of
the Sky"\cite{pi_of}.  All transients are highly variable and therefore 
snapshots must be taken at intervals of several seconds or even less. Hence solutions capable of operating 
without dead time (such as the time needed for ccd readout), like MASTER VWF \cite {master}, TORTORA and others  (see
table~\ref{tab_wf}), have an important advantage over other systems.

The operation of any automatic astronomical facility is supported by a dedicated software package, which 
also performs primary image processing and data storage. Which software to use for an automatic astronomical 
system depends on the particular task the facility is meant to perform, and  multi-purpose solutions are therefore
unsuitable in such cases. Classical universal techniques are even less suited for super wide field observations,
because of the problems due to the field curvature and very large data  amount.

\begin{figure}[!t]	
\psfig{figure=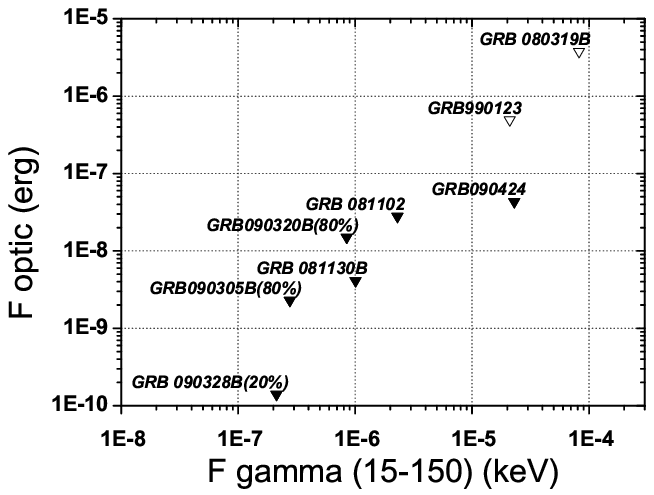,width=78mm}
\psfig{figure=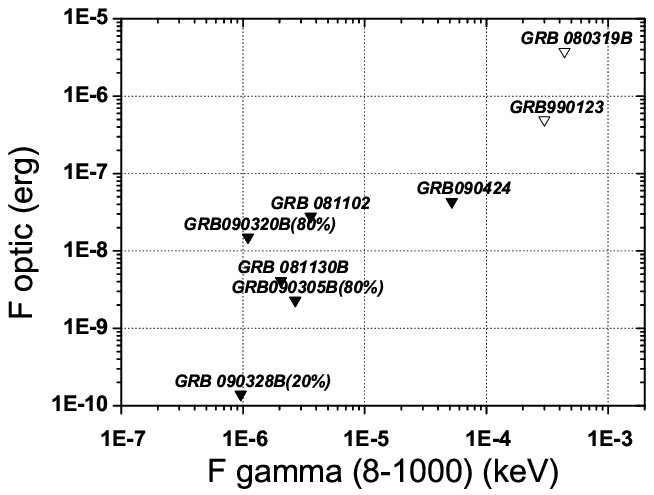,width=78mm}	
\caption{ Summary results of synchronous observations of a gamma-ray burst.
Dependence plotted as a function of gamma-ray fluence in the soft ($15-150 keV$) and hard ($8-1000keV$) 
energy interval.  }
\label{allgrb}
\end{figure}

\subsection {MASTER - Mobile Astronomical System of the TElescope Robot}

MASTER was the first robotic telescope of this kind constructed in Russia. It was built and
tested by students, post-graduate students and lecturers of SAI MSU in 2002
(http://observ.pereplet.ru). MASTER has evolved into a broad network of
telescopes spread across the entire country. A detailed description of the MASTER facility can
be found in~\cite{lipunovmaster}. Below we discuss the basic properties of
MASTER wield field cameras (MASTER-VWF).

The main task of the MASTER-VWF experiment is to perform continuous all-sky monitoring with the aim to
detect all objects absent in the available astronomical catalogues. This includes, in
particular: 
\begin{itemize}
\item  Prompt detection of grb emission synchronously with gamma-ray space
observatories.
\item Search for star-like transients of unknown origin and orphan bursts. 
\item Detection of meteors and determination of their basic parameters:
brightness (photometry), velocity (astrometry), and altitude of combustion in the
atmosphere (triangulation).
\item Detection of satellites and space debris and determination of their basic parameters
- astrometry, velocity, brightness, and altitude above the earth's surface.
\end{itemize}

\begin{figure}[t]
\psfig{figure=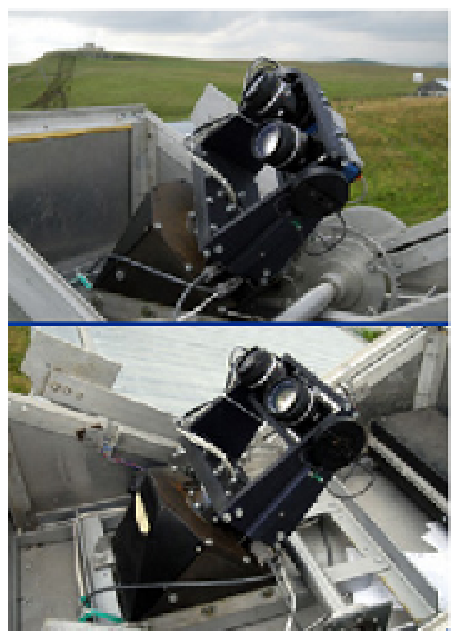,width=52mm}
\psfig{figure=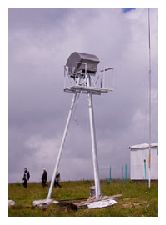,width=52mm}
\psfig{figure=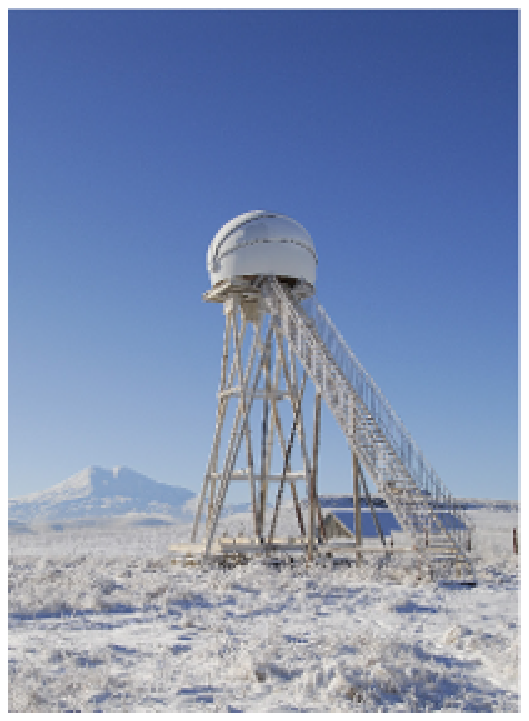,width=52mm}
\caption{The equipment of the MASTER facility deployed at the MSU
Caucasian Mountain Astronomical Observatory  near Kislovodsk. From the left to
right: (1) the East and West cameras of the  MASTER-VWF4 northern installation; (2) the
main tower and shelter of the southern MASTER-VWF4 installation, and (3) the main tower
of the MASTER-2 telescope \cite{lipunovmaster} }
\label{mas_all}
\end{figure}

For these purposes, a completely robotic MASTER VWF4 facility with four wide-field
cameras and a total field of view of 4000 square degrees has been deployed at the MSU Caucasian
Mountain Astronomical Observatory   near Kislovodsk(2075m). Another facility ---
MASTER VWF2 (two cameras and 2000 square degrees field of view) --- has been deployed
near Irkutsk. In Kislovodsk cameras are set on parallactic mounts in
pairs  702 meters apart (see fig.~\ref {mas_all}). Each mount has an automated dome and 
two fast and powerful (11 Megapixels) Prosilica GE 4000 CCD cameras  with  Nikkor  50mm ($f/1.2$) lenses.
The camera is capable of continuous imaging with a speed of one frame in  0.2 to 60
seconds with no time gaps.  The 702-m baseline allows the facility in Kislovodsk
to determine the parallaxes of circumterraneous objects and reconstruct their altitudes
from the results of astrometric reduction.

\begin{figure}[!t] 
\psfig{figure=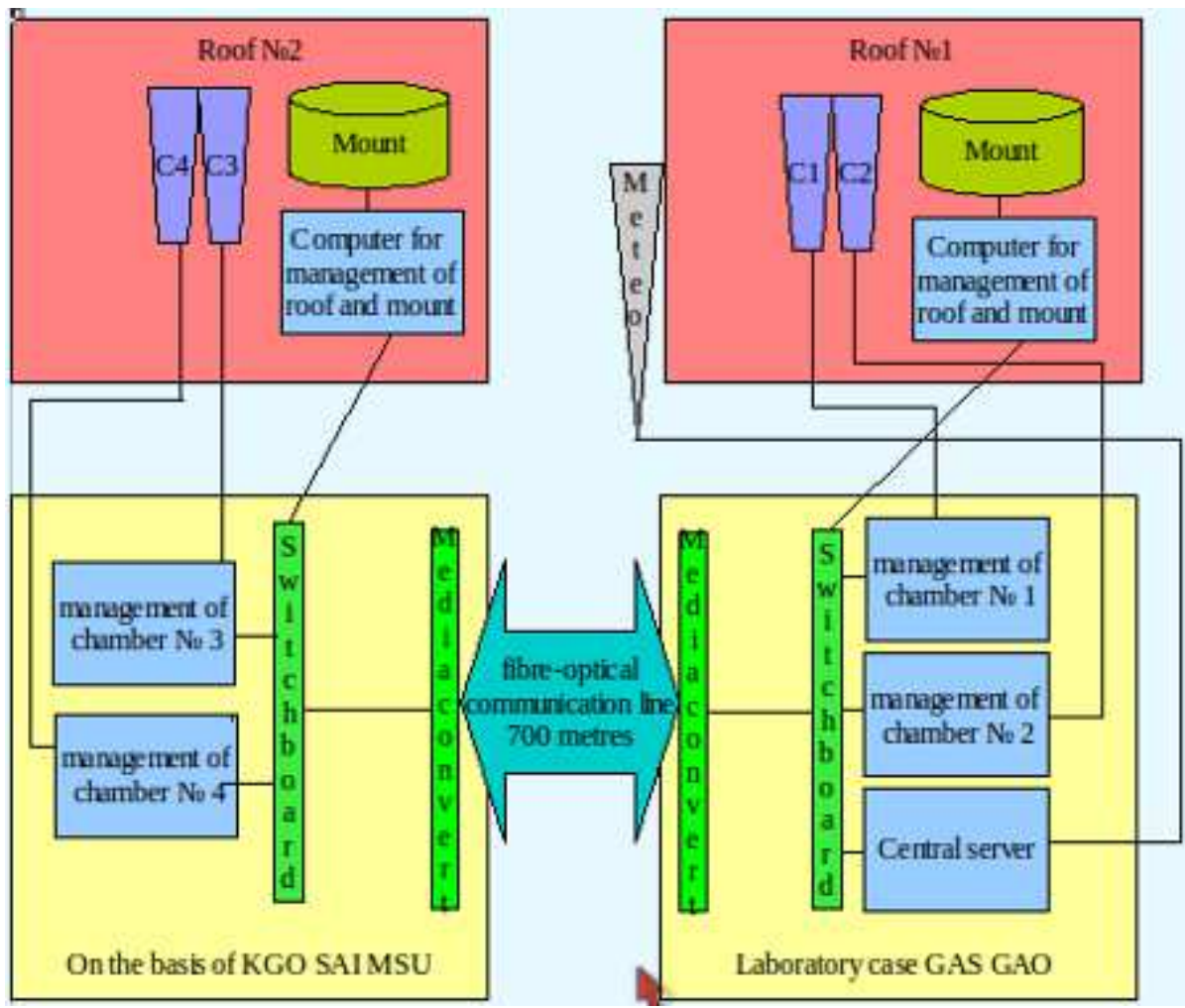,width=160mm}
\caption{Schematic diagram of the layout of the equipment in the MASTER VWF4 experiment.  }
\label{vwf_scem}
\end{figure}

All telescopes operate in a fully autonomous mode: at night, if the weather conditions permit
(a "Boltwood Cloud Sensor" provides the current weather report updated every three seconds), the
cameras begin to automatically survey the sky. The frames taken are then  processed and stored in 
a special data storage device. Because of the very large data flow (up to 700 Gb per night), the images
can be kept only for several days and then most of them have to be erased, except for some images with objects 
of special interest like  meteors, candidate optical transients, and satellites. We also keep the images taken
simultaneously with grbs. The survey is halted and the telescope dome is closed every time when the weather
deteriorates  
or at dawn. The operation of the entire MASTER VWF4 facility is controlled by seven computers 
(see Fig. \ref {vwf_scem} ):
four computers to control the operation of the  Prosilica GE 4000 CCD camera and perform source extraction;
two computers to control the operation of the dome and mount;
one big quad-core server to process and store the data obtained and publish it in the Web.

The facility is a complex engineering system, which needs sophisticated and powerful software 
to control it, perform observations, and data processing. The data flow can be as high as several 
Terabytes/day (in the case of very short exposures on the order of 0.2s), requiring highly optimized
software, powerful data storage facilities, and interfaces. With 5-second long exposures the limiting
is 11.5-12 on each frame depending on weather conditions. Thus every image contains about 10-15 thousand objects,
resulting in a data flow that is impossible to store and has to bee processed in real time.

\begin{figure}[!t] 
\psfig{figure=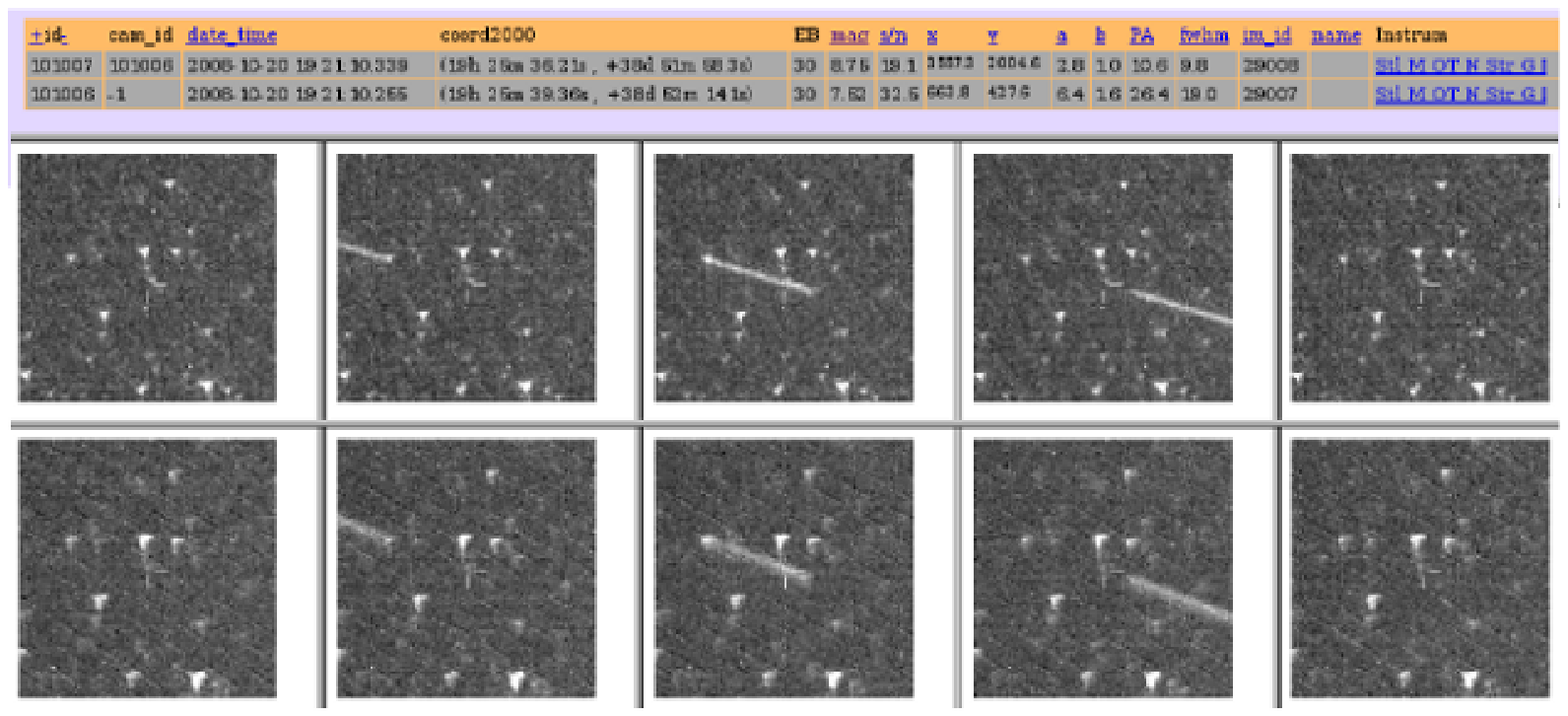,width=160mm}
\caption{Debris of the American military satellite USA114Deb found via the web interface of the MASTER system.
The top and bottom rows of the images are taken by the Northern and Southern cameras, respectively. 
The parallactic shift is apparent event to an unaided eye. The measured parallax implies a height of  
$4000 \pm 200$ kilometers}
\label{usa114}
\end{figure}

A special software package has been developed for this purpose. It allows the
coordinates of all objects and their photometric characteristics (profile, magnitude) 
to be determined in real time, and is also capable of finding and analysing optical
transients. MASTER VWF can measure the coordinates of the objects with an error
smaller than 10 arcsec. The system is supplied with a daily updated database of the 
ephemerides of artificial satellites and a program developed to identify the satellites in 
real time. The program is capable of finding new artificial satellites, to observe the destruction
of known artificial satellites and not only to control space debris, but also
observe the process of its formation. Sometimes unknown or lost objects are recorded. 
For example, on October, 20th, 2008 MASTER VWF recorded the debris of the
American military satellite USA114Deb  (see fig. ~\ref {usa114}).

A total of six synchronous observations of gamma-ray bursts have been made by  MASTER VWF in Kislovodsk and Irkutsk
during seven months of their operation.  In all cases high upper limits
have been determined (see tabl ~\ref {tab_grbwf} and fig. ~\ref {allgrb}).

In the following chapters we describe these and other results in more detail.

\section {Primary reduction in wide fields: astrometry and photometry}
Let us consider the methods of astromeric and photometric reduction of a wide field in more detail.
The identification of the field of view and calculation of the coordinate grid for the
image is based on the table of Cartesian image-plane coordinates (X,Y) 
produced  by program used for the extraction of objects (SExtractor) and the the Tycho-2 astrometric catalogue,
which contains exact coordinates for most of the stars down to a limiting magnitude of $11^m.5$. To perform this task,
the program should meet high requirements in terms of speed of processing (10-15 seconds and less) for
a field containing about 10000 stars; ensure high probability of 
identification (more 99 \%); be applicable to various fields ranging in size from
from narrow (tens of square minutes) to super-wide (thousand square degrees) fields; allow for
a certain  freedom in the telescope parameters, because the focal length and
other parameters may change in the course of the year due to the variation of the 
outside temperature, and be capable of operating with approximate initial midfield coordinates
provided that they are within the field of view.

\begin{figure}[!t]
\psfig{figure=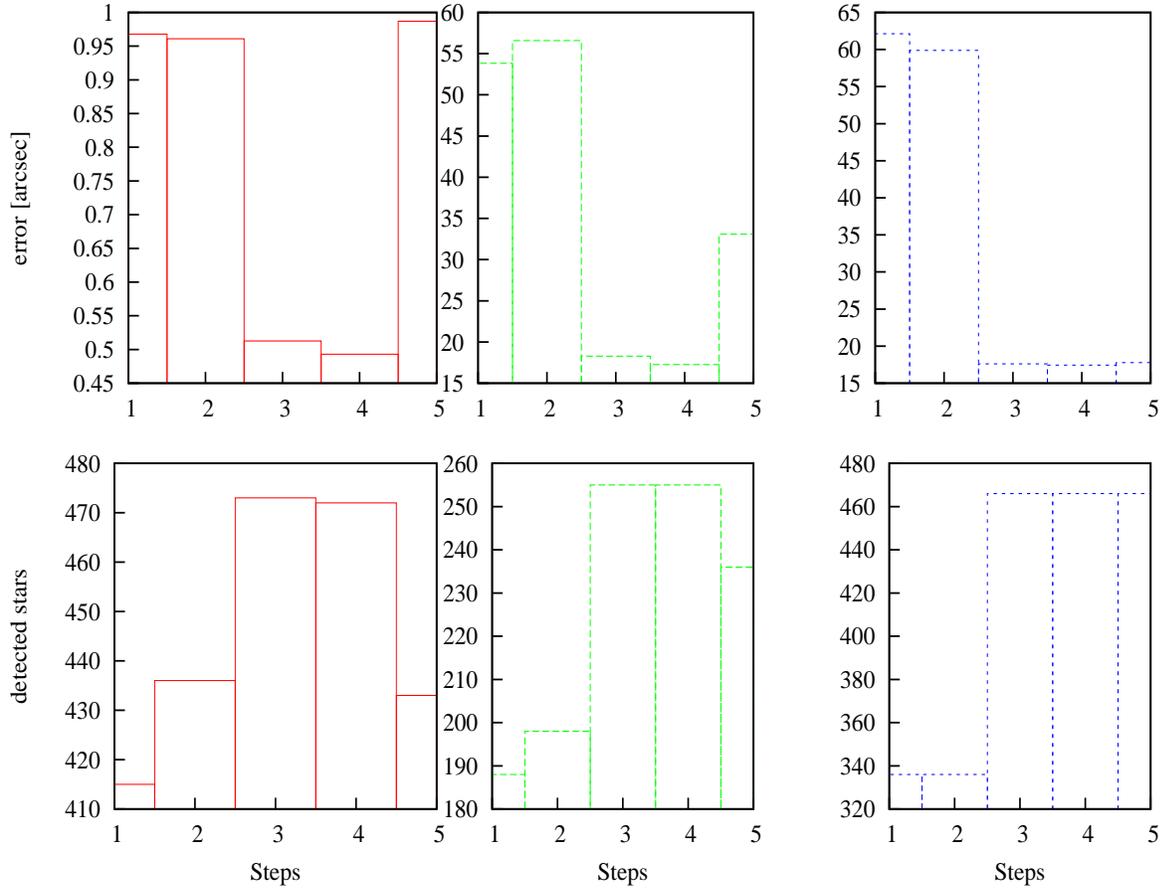,width=160mm,height=120mm}
\label{astrom_6}
\caption{ Average error (the upper curve) and the number of identified stars (the bottom curve) 
as functions of the degree of the polynomial used. The dependences
are shown for three different optical systems. Red (left) Rigter-Slefogt
(MASTER) with a field of view of $6 {_ \square} ^o $,  Nikkor 50 mm f/1.2 (middle) with a
field of view of $1000 {_ \square} ^o $, and Nikkor 50 mm f/1.4 (right) with a field of
view of $1000 {_ \square} ^o $. The curves are based on arbitrary frames from various sky areas.
A third to fourth degree polynomial approximation can be seen to provide optimum solution
(95-99 percent of stars are identified with the minimum average error). A more detailed analysis 
shows that the third-degree fit yields slightly better results. That is logical, because most of the known
aberrations (e.g., distortion) depend on the third power of radial distance.}

\end{figure}

All processing in  MASTER experiments is performed under Linux OS and hence
no ready-made software was likely to be found. The current implementation of the "astrom" 
program is based on the algorithm of pairing similar triangles of bright
stars in the image and in the catalogue. This program works perfectly with the most
different images from those taken under poor weather conditions and containing
a few dozen stars (typical situation for alert observations)  to images in the
Milky-Way fields taken during a good observing night and containing up to 100000 objects.
The root-mean-square error of the measured coordinates ranges from 
$ ^1 / _ 4$ of arcsecond (scale 2.1"/pixel)  for most of the images taken with the basic 
telescope of the MASTER system to about 7-8 arcseconds for very wide-field cameras
(scale 36"/pixel) (see fig. \ref {astrom_6}).

The system may also operate in  a ``fast'' astrometry mode to process consecutive frames of 
the same area taken with wide-field cameras. In this mode the software uses a priori information about
the field of view based on the previous frames, allowing the operating time to be reduced to less
than 0.2 seconds. There is also a dedicated mode for more accurate determination of the coordinates of 
special objects (gamma-ray bursts, supernovae, asteroids, meteors, and others)
with the local coordinate adjustment computed only in the field of the object searched 
using 10-25 nearest stars with well accurate coordinates. Our experience shows that 
this mode provides a factor of two improvement of the  accuracy  of the inferred coordinates.

Our program for extracting objects from images is based on the SExtractor  software package 
written for the TERAPIX \cite {sext} project. 
SExtractor is written in the C language and offers sufficiently high performance. We adapted this 
program to the task of reducing images taken with MASTER, and describe our version of the program 
in the following chapter. 
The main parameter used by the extraction program is the threshold 
signal-to-noise ratio. This value directly affects both the program execution time and the
accuracy of the results obtained. We adopt S/N=2 for primary selection, whereas at the
subsequent stages of processing we select from the initial list only the objects with higher S/N 
values.

\subsection {Photometric calibration}
		
\begin{figure}[!t]	

	\psfig{figure=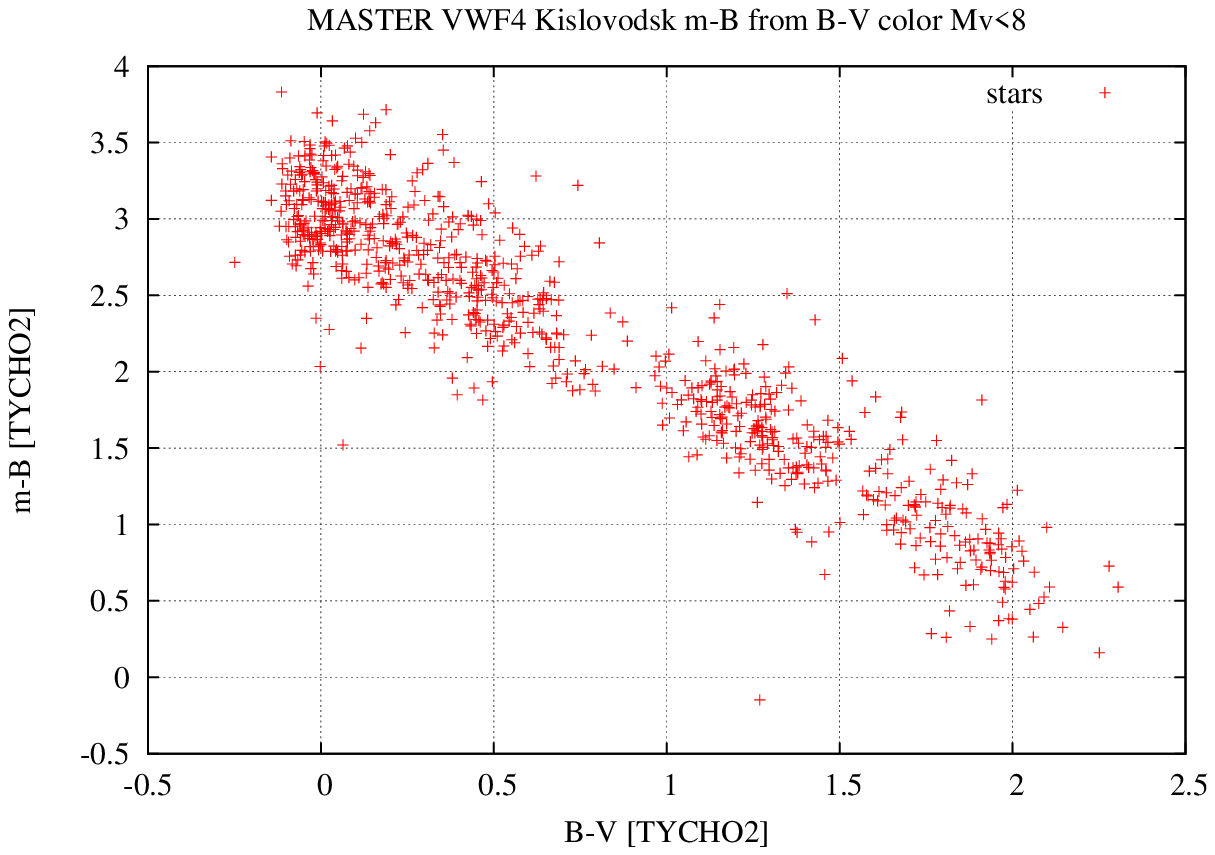,width=78mm,height=70mm}
	\psfig{figure=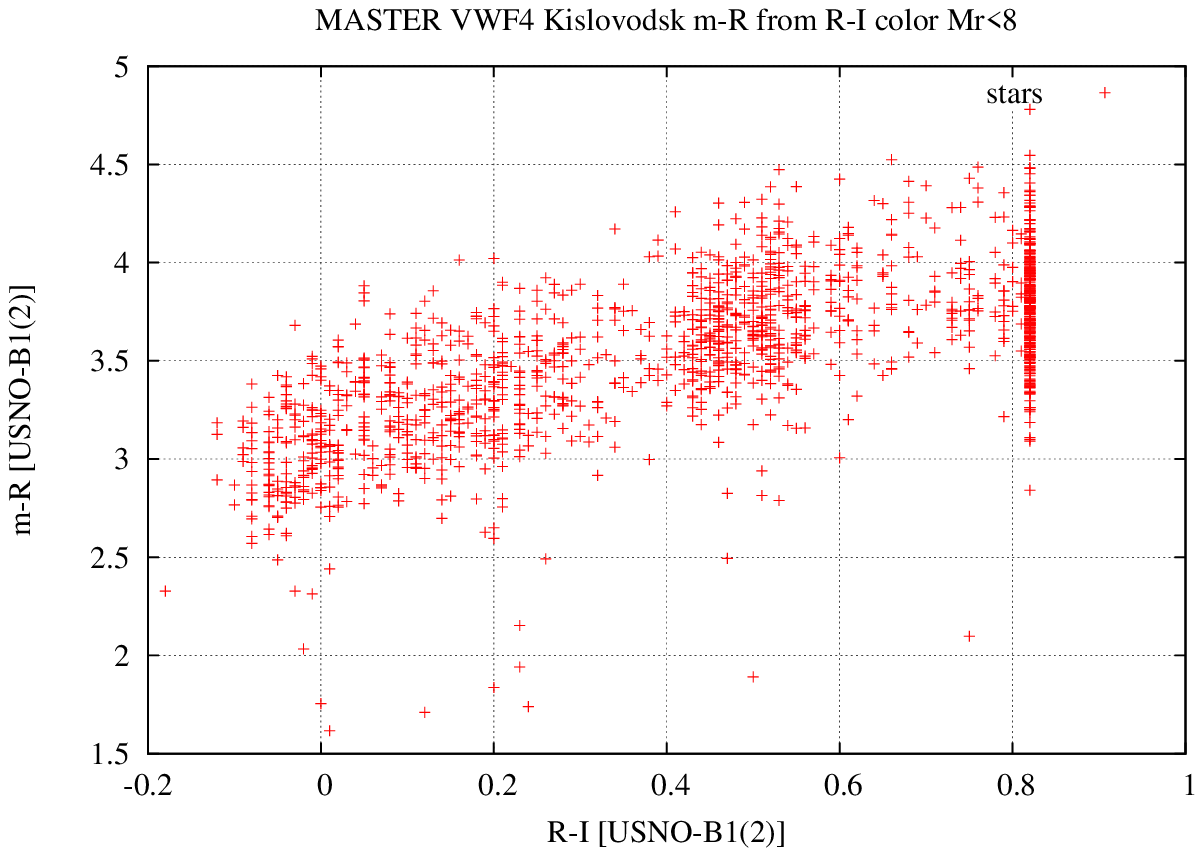,width=78mm,height=70mm}
	\psfig{figure=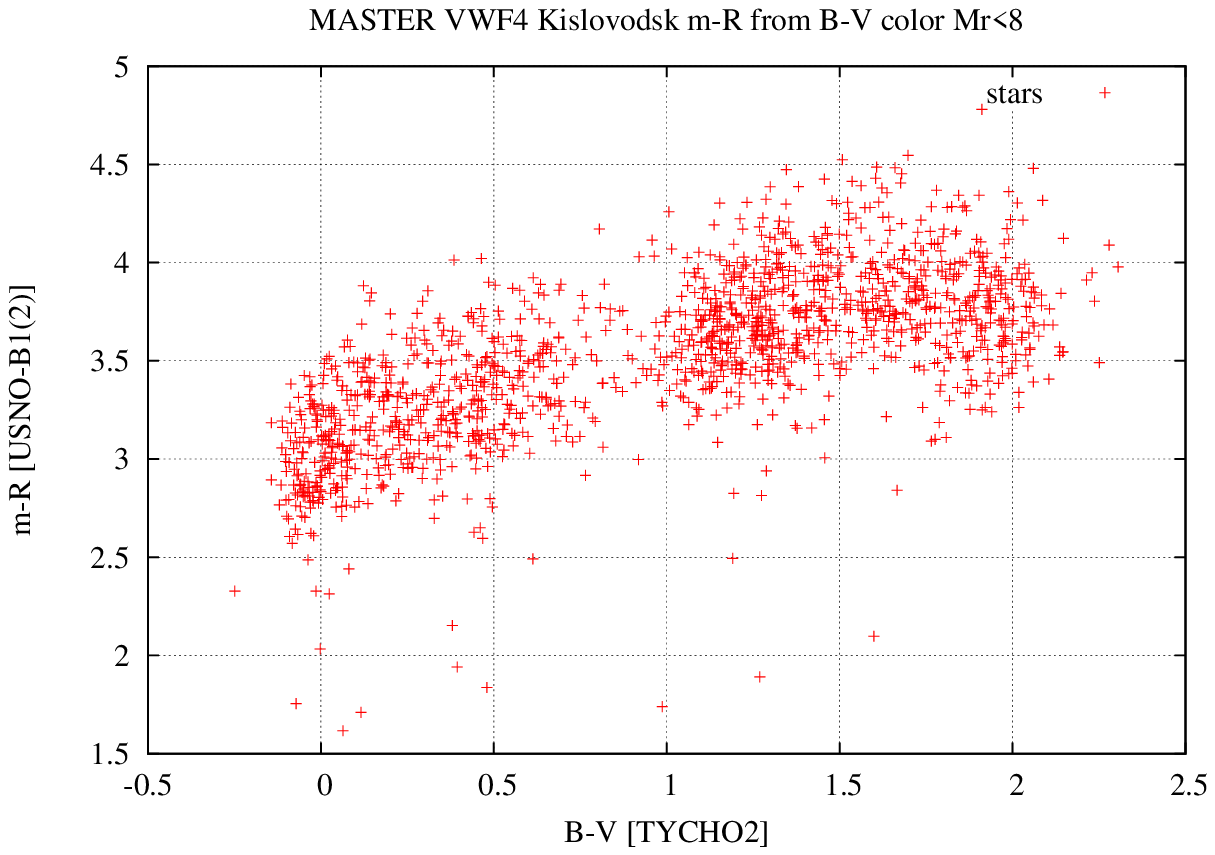,width=78mm,height=70mm}
	\psfig{figure=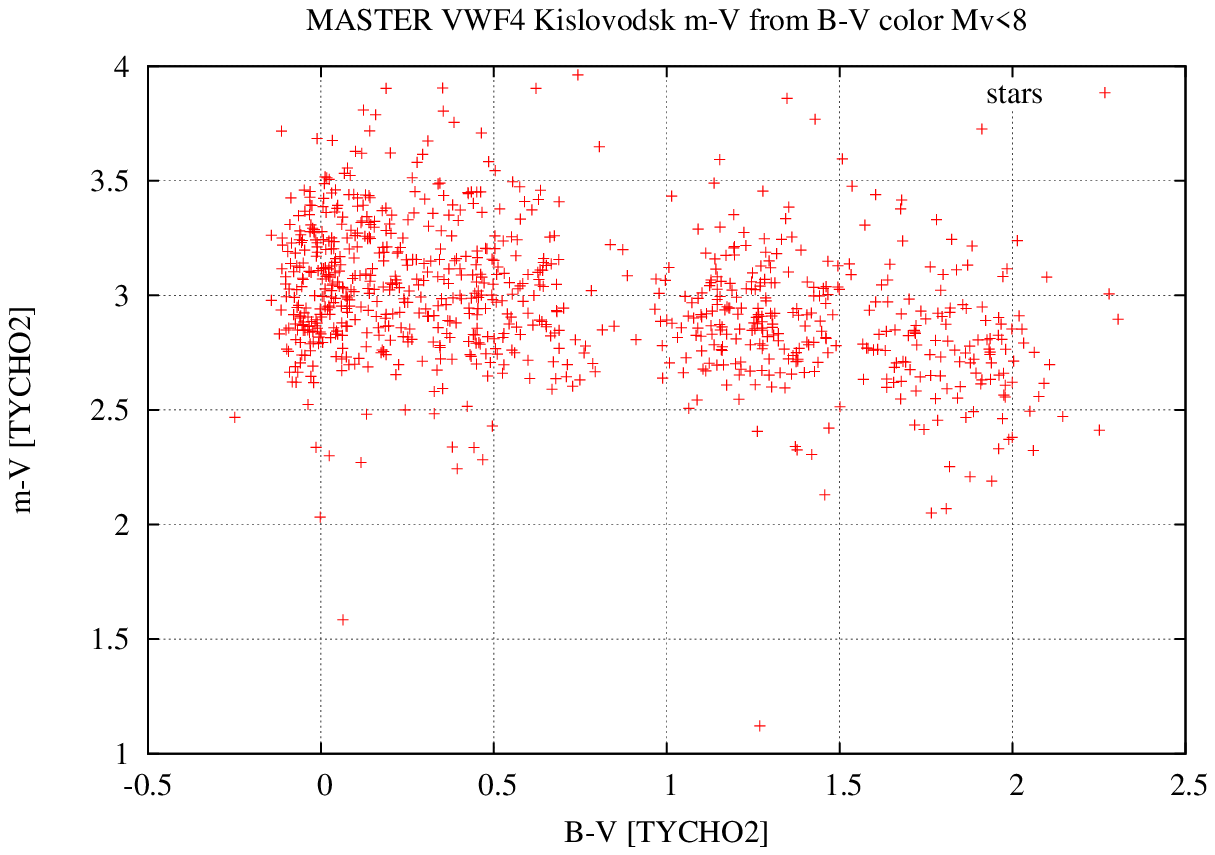,width=78mm,height=70mm}	
	\caption{ 
colour-colour diagrams are used to calibrate  instrumental magnitudes.
On these plots $m=2.5lg (F _ {inst}) $ and the zero point is of no importance and cannot be 
determined at the this stage. Note that the width of the sequences is real and due to the fact
that stars have different spectral types, and not due to measurement errors.
}
	\label{photcalib}
	\end{figure}
	
Practically all wide-field systems operate without filters. MASTER VWF is not an
exception. This poses the problem of the calibration of the photometric band of the measurements.
It is clear that our instrumental photometric band is 
limited to a certain extent by the CCD sensitivity curve  and lens transmission curve. 
Our task is therefore to determine  instrumental magnitudes based on the magnitudes of the
reference stars adopted from  photometric catalogues.

To this end, colour-colour diagrams are analysed, which relate the difference between the instrumental magnitude 
and the magnitude in any of the modelled bands (or combinations thereof) to a chosen colour index for each 
detected star. The closer is the modelled band to the instrumental band, the more horizontal appears the
the cloud of points on the diagram. Fig. ~\ref {photcalib} shows the
corresponding colour-colour diagrams for the MASTER VWF-4 system consisting (in its photometric part)
of a Nikkor 50 mm f/1.4 lens and  Prosilica GE4000 CCD. The instrumental photometric band can be described fairly well
by the $V _{TYCHO2} $ filter. The absolute accuracy of photometric measurements made with
wide-field cameras is, on the average, of about $ \Delta m _ {abs} \sim 0.25-0.35^m $. Note that the
relative accuracy of measurements is substantially better with errors as low as $\Delta m _ {rel} ~ 3-5 \% $.

\subsection {Automatic determination of the limiting magnitude of an image}

\begin{figure}[t]
\psfig{figure=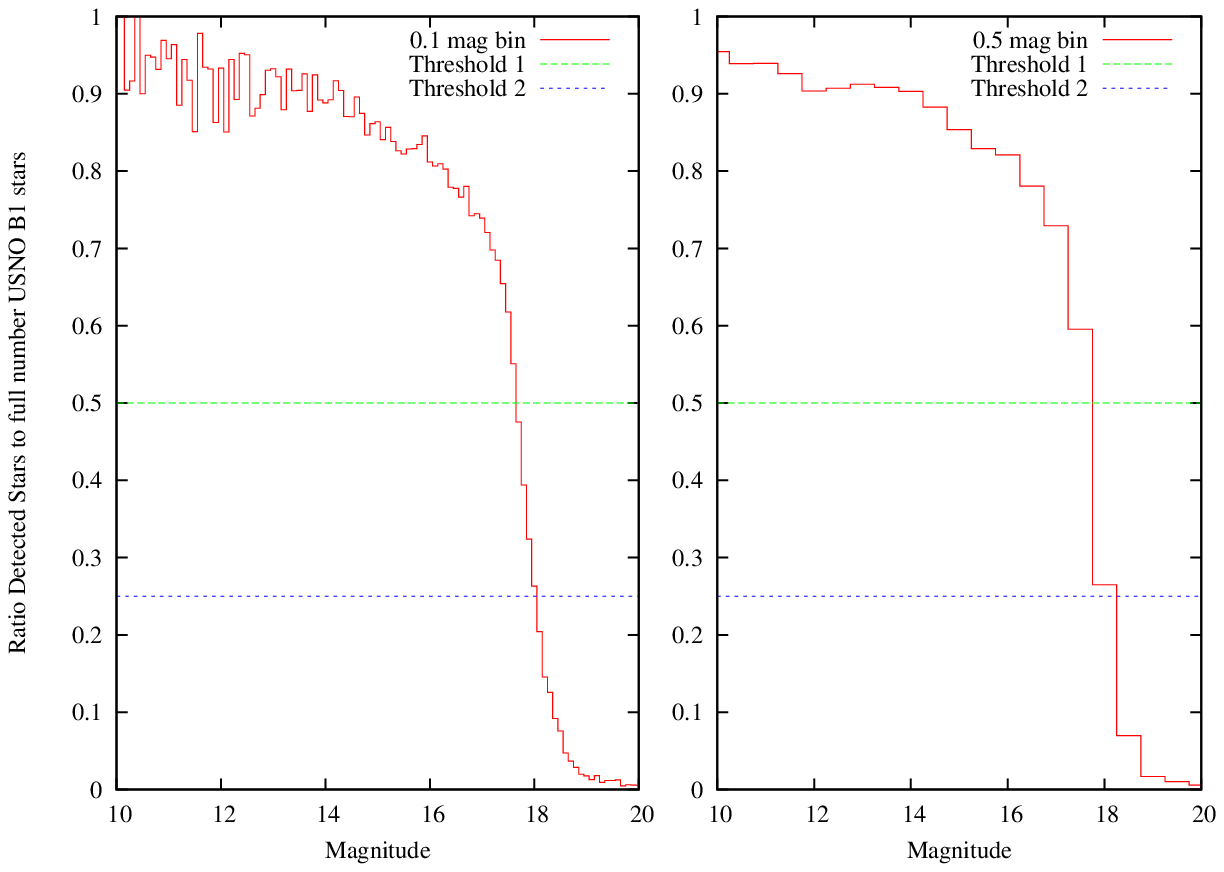,width=160mm,height=80mm}	
\caption{The number of identified stars as a function of magnitude. The straight
lines indicate the $ \sigma _ {lim} =50 \% $ and $ \sigma _ {lim} =25 \% $ thresholds.
Averaging was made within $0.5^m $ and $0.1^m $ bins in the right and left histograms,
respectively. The right (coarser binned) histogram is smoother,
resulting in reduced probability of accidentally reaching the threshold at a bright magnitude. 
It is therefore convenient to determine the limit on the coarsely binned histogram and
then refine it on the finer-binned curve.}
	\label{flimits}
	\end{figure}

Automatic determination of the limiting magnitude for the images taken is an important
part of the operation of modern automatic sky surveys. An automatic survey must naturally
include automated quality control. The automatically determined limiting magnitude may
serve as some kind of objective quality assessment for the images obtained. 
Furthermore, the determination of the limiting magnitude is also needed 
for gamma-ray burst observations.

The task of the determination of the limiting magnitude can be addressed formally by setting
the limiting magnitude of the frame equal to the magnitude of the
faintest object. This approach, however, has a number of shortcomings.  First, when the 
system is operated with a low extraction threshold most of the faintest objects are noise 
artifacts. If the sky is covered with clouds, no real faint objects can be seen and practically
all objects at the visibility limit are chance fluctuations of cloud brightness. Moreover,
because of the large number of such false objects some of them may be identified with
real stars and hence the situation cannot be remedied by restricting the list to objects
identified with catalogued stars. Second, this approach allows one to estimate the limiting magnitude
only within a local area and the quantity inferred does not characterise the entire image and is 
therefore unsuitable for quality control.

These problems can be addressed as follows. All stars of the reference catalogue that are located within the
field considered are subdivided into magnitude bins of fixed width. For each bin the fraction of 
identified stars to the total number of catalogued stars is computed, and the corresponding
histogram is drawn  (see fig. ~\ref {flimits}). One can see that the ratio decreases sharply when
the average magnitude of the bin approaches the limiting magnitude. In practice, it is convenient to 
adopt a certain threshold ratio and define the limiting magnitude as the magnitude at which the 
curve reaches this threshold. The thresholds of $ \sigma _ {lim} =25 \% $ and $ \sigma _ {lim} =50 \% $ 
have been shown empirically to correspond 
to  objects with $^S / _ N \approx 9-11\sigma $ and $ ^S / _ N \approx 14-15\sigma $,
respectively.

Note that for robust operation of the algorithm the limiting magnitude of the reference catalogue
should be fainter than that of the frame studied. Otherwise the method becomes inapplicable because 
the catalogue lacks lack stars of the given magnitude. However, the USNO-B1 catalogue is sufficient
for most of the modern wide-field surveys, whereas very wide-field surveys can be reduced successfully even 
with the TYCHO-2 catalogue (and this is very convenient).

\section {Automatic classification and analysis of astronomical transients.}

\subsection {Extraction of objects and data transfer to the server.}

The search for and classification of transients starts once the primary reduction of
the image is finished. All transient events in frames can be subdivided into two classes:
starlike objects and strips, which differ substantially in the methods used for primary 
reduction. All  rapidly moving objects are usually classified as strips: 
low-altitude satellites and meteors. All transients of stellar or extragalactic nature (novas and
supernovas, orphan bursts, stellar flares) and geostationary satellites
(see, for example, fig.\ref {tr_grb}) are  starlike. Starlike objects are much
easier to process, because the program of object extraction  and 
instrumental photometry (SExtractor, described above) has been developed to analyse such objects.
Therefore all main parameters of starlike objects (flux and magnitude, FWHM,  Cartesian
frame coordinates (X, Y), and semiaxes (a, b)) are determined very reliably. 

\begin{figure}[!t]
\psfig{figure=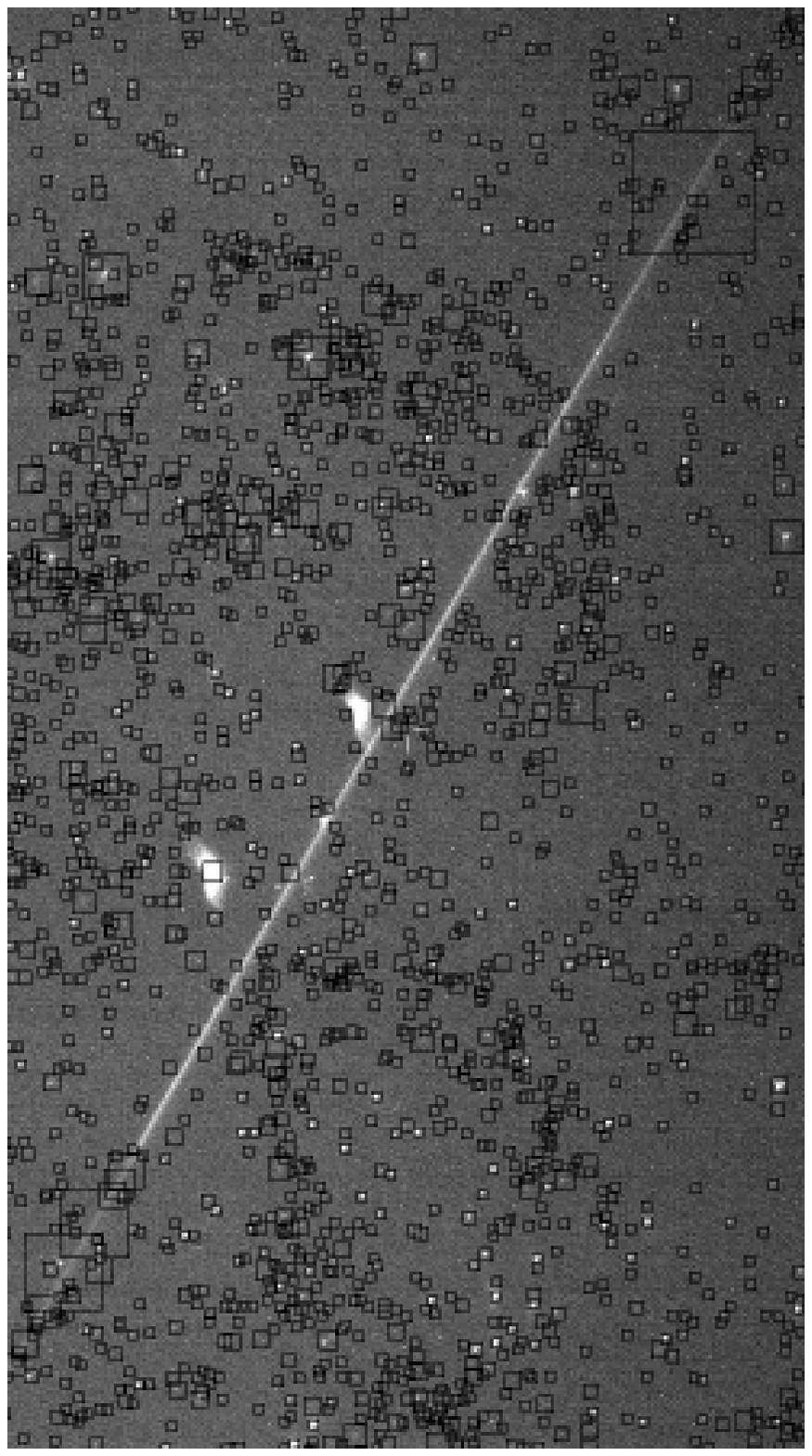 ,width=79mm,height=120mm}
\psfig{figure=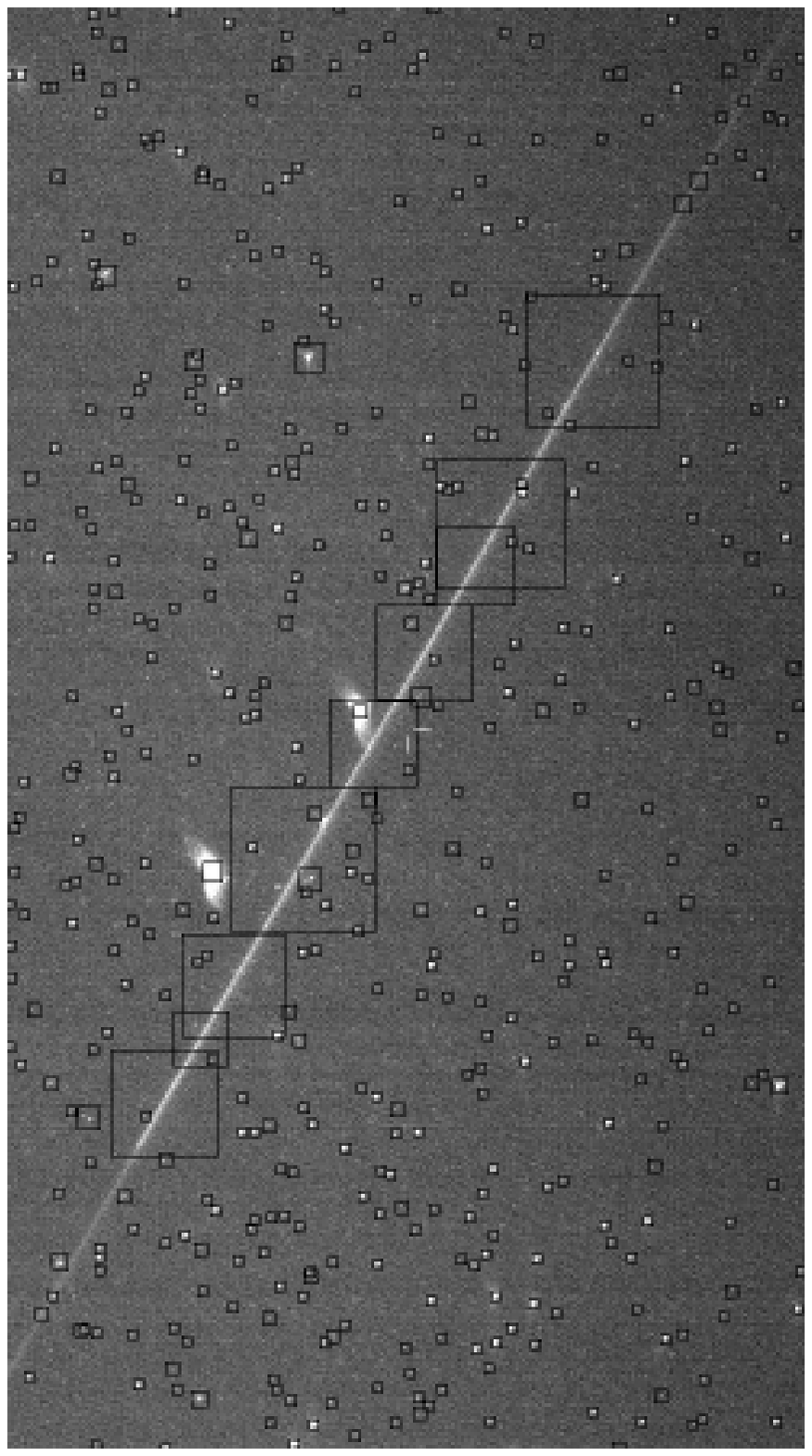,width=79mm,height=120mm}
\caption{A  5 $\circ \times 9 \circ$ image with a bolide. The black
squares indicate  objects extracted from the frame. The side of the square
corresponds to the fwhm of the object. The left-hand figure shows a frame with objects extracted
using standard SExtractor parameters (with the extraction threshold of $ ^S / _ N> 1.5$) with no filters 
applied. Ax a result, the program identified all starlike objects, heaps of faint garbage, hot pixels, and
simply random noise, i.e., virtually everything except the bolide. The right-hand frame shows the result of
extraction made using the currently adopted refined parameters. Several bright elongated objects are extracted
on the bolide image and the noise level is appreciably lower.}
\label{met_videl}
\end{figure}

Extraction of strip-shaped objects is fraught with serious problems. First, a fine adjustment
of SExtractor's additional internal parameters  is necessary. Second, a change of
these parameters strongly affects the image processing time. Furthermore, 
strip-shaped objects contribute appreciably to the background within the small
area suitable for identification of stars. Fine adjustment of all parameters
allowed us nevertheless not only to identify strips in the image, but also achieve high speed of operation
(half exposure time)  (see fig.\ref {met_videl}). However, SExtractor algorithm cannot identify strips in the frame, 
to say nothing about performing their high-quality photometry. To this end, a special program
had to be developed, which we describe below.

To reduce the server load, identification of objects in the image can be physically performed not only by the
server, but also by the control computer of each particular camera (see ~ fig. ~\ref {vwf_scem}). 
Note that the resulting catalogue of detected objects should have higher priority for the transmission to the server
compared to the images that are to be stored in the database.

\subsection {Primary reduction of astronomical transients.}

The next stage after identification of objects is their astrometric and photometric reduction, which we 
describe in detail in the previous section. We thus obtain, after identification, astrometric and
photometric calibration, a catalogue of objects identified in the frame together with calibrated coordinates
and photometry. All unidentified objects (identification is performed using both coordinate
and photometric criteria: usually $\Delta S=20-30 "$ and $ \Delta m=1.5^m $) are viewed as
possible optical transients. This approach allows us to identify stellar flares as
transients and to a certain extent prevent spurious identifications with
fainter and hence more numerous stars. To filter noise, objects that could not be
identified with the catalogue are correlated with the objects of several (usually
2-3) previous sets. An object is considered to be a candidate transient if it could not
be found either in the catalogue or in in the previous frames. All candidate transients are then
passed through a number of special filters. For example, if the semimajor axis of the object, which 
should typically be of about $b \sim 0.7-2$ pix, exceeds $b \sim 4-5$ pixels, it must be
a trail of a cloud with a probability of 95 \%. If $fwhm<1$ pixel, we must be dealing with a hot pixel or 
a cosmic-ray hit. We also set for all transients the trigger threshold, which is always higher than the
extraction threshold.  Currently, the trigger threshold is $ ^S / _N> 10 \sigma $. We consider a candidate found on 
two or more consecutive frames to be a boda fide transient.

\begin{figure}[t]
\psfig{figure=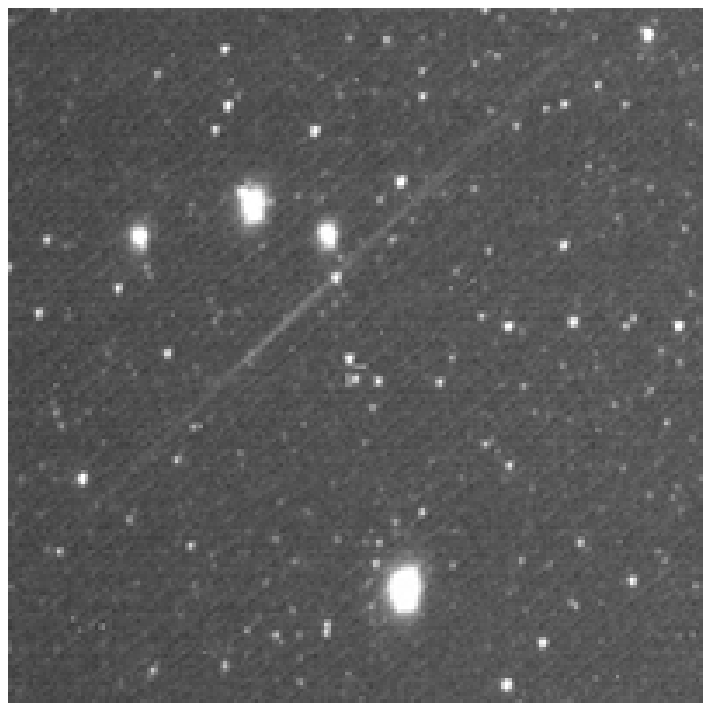,width=80mm}
\psfig{figure=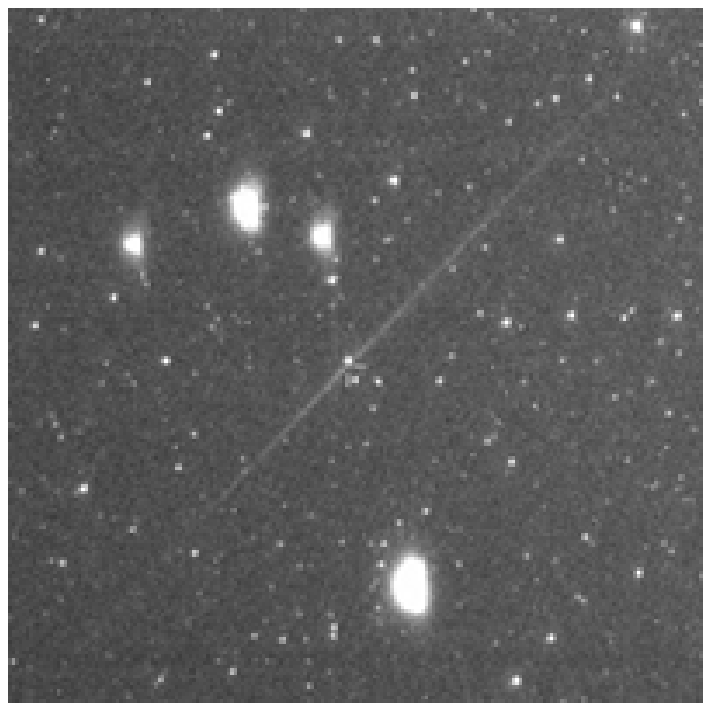,width=80mm}
\caption{  MASTER VWF4 frames of a $5 ^ {\circ} \times 5 ^ {\circ} $ sky area taken with the Northern and Southern
cameras. The parallax of the meteor is immediately apparent. Triangulation yields a height of
$H _ {meteor} =72 \pm of 2$ km. }
\label{met_par}
\end{figure}

Preliminary classification of transients is performed after the objects are passed through all filters. 
First, all objects are searched for  possible identifications in the catalogue of artificial 
satellites and all identified objects are included into the database of satellites.
The remaining ones are subdivided into starlike and strip-shaped  objects. The object is classified
as a strip-shaped if $a/b>3$, otherwise it is considered to be starlike (here $a$ and $b$ major and minor semiaxis). 
All starlike objects are then passed through the coincidence circuit, and strip-shaped objects are. Strips are
provisionally
subdivided into meteor candidates and unknown artificial satellites. If each of the three successive frames
contains a candidate transient such that the three objects lie on a straight line, all of the them are considered
to be the images of an artificial satellite. The remaining objects are considered to be meteor candidates.

\subsection {Coincidence scheme of  and determination of heights.}

\begin{figure}[t]
\psfig{figure=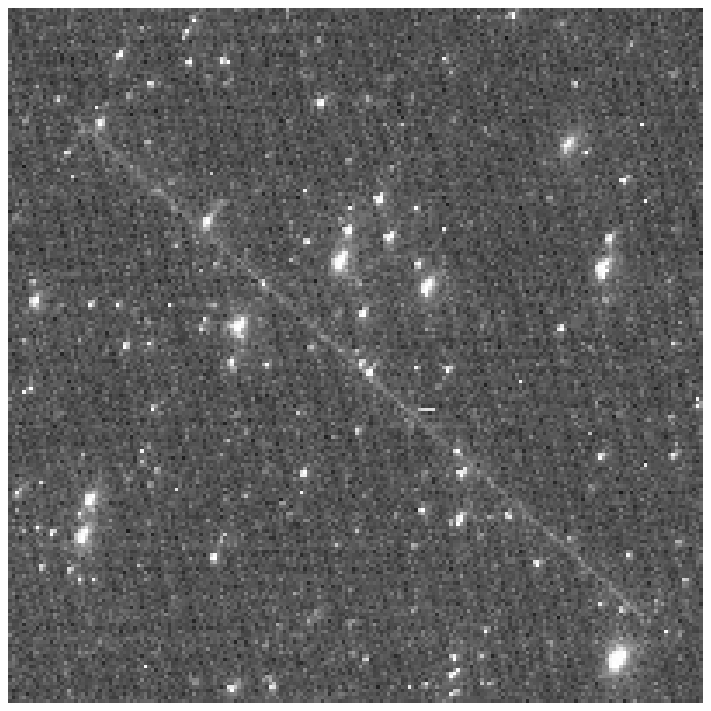,width=80mm}
\psfig{figure=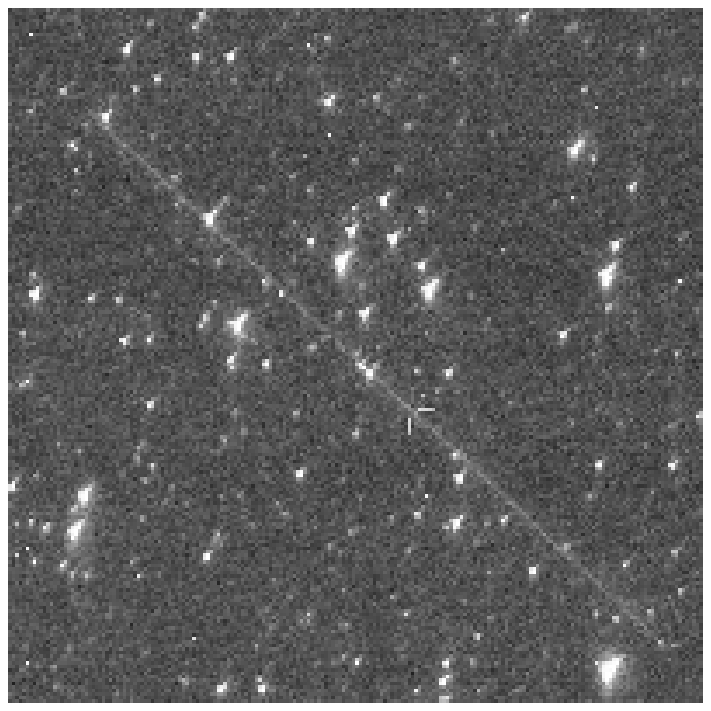,width=80mm}
\caption{ Parallax of an artificial satellite. Here we show the  $4 ^\circ \times 4 ^\circ $ frames
of a sky area taken with the MASTER VWF4 Northern and Southern cameras. Triangulation yields a height of
$H _{satell} \sim 4500$ km. Given the accuracy of astrometric measurements for strip-shaped objects,
$ \sigma _ {astrom}^ {line} \sim 20 "$, and the baselength of $ \Delta l=702$~m, the maximum height 
that  can be determined via triangulation is $H _ {max} = \Delta l*arcsin (\sigma _ {astrom} ^ {line}) \sim 10000$m. }
\label{sat_par}
\end{figure}

All the procedures described above are performed for each of the four cameras in real time mode. 
Real time processing ends here if the cameras operate in uncoupled (alert) mode. In the case of
survey observations, when two cameras point to the same sky area, all starlike transients 
are passed through the coincidence scheme, and all strip-shaped objects have their heights
determined (see fig. \ref {met_par}, \ref {sat_par}). To improve the accuracy of identification,
special transformation is applied to the frame taken with the Northern camera to convert it to the
coordinate frame of the Southern camera. This transformation is based on the same principles
as astrometric reduction. It takes  little time to compute
($\sim $ 0.2 seconds) and improves the accuracy (and hence reduces the correlation radius and
the number of chance identifications) by more than a factor of 2.5 compared to the accuracy of the results
obtained in the case of identification by celestial coordinates. Should a real starlike transient be found,
an alert can be immediately issued to larger (45-cm) MASTER telescopes (located near Moscow, 
in Kislovodsk and in the Urals) to study the event in more detail. 
However, this mode is currently in the stage of development, mostly because of the debris of
high-orbital satellites, which often contaminate the frames (see
fig. \ref {usa114}, \ref {tr_grb}).

\subsection {Analysis of strips.}

As we pointed out above, SExtractor computes only the approximate position of a
strip, breaking it into several bright elongated objects. To perform  complete
photometry of a strip, determine the coordinates of its centre and ends
and other parameters, a special program is used dedicated to the analysis of strips
on astronomical images. To prevent server overload, this program is run only 
when cameras are idle (in the afternoon or during bad weather).
The program starts with the coordinates of any location inside the
strip determined at the previous stage. The working area of the given size 
(usually 512 on 512 pix, which is sufficient for the analyses of any satellite or
meteor trails) is chosen around this initial point. The size of the working area is 
automatically increased when a very bright and long bolide is to be analysed.

\begin{figure}[!t]
\psfig{figure=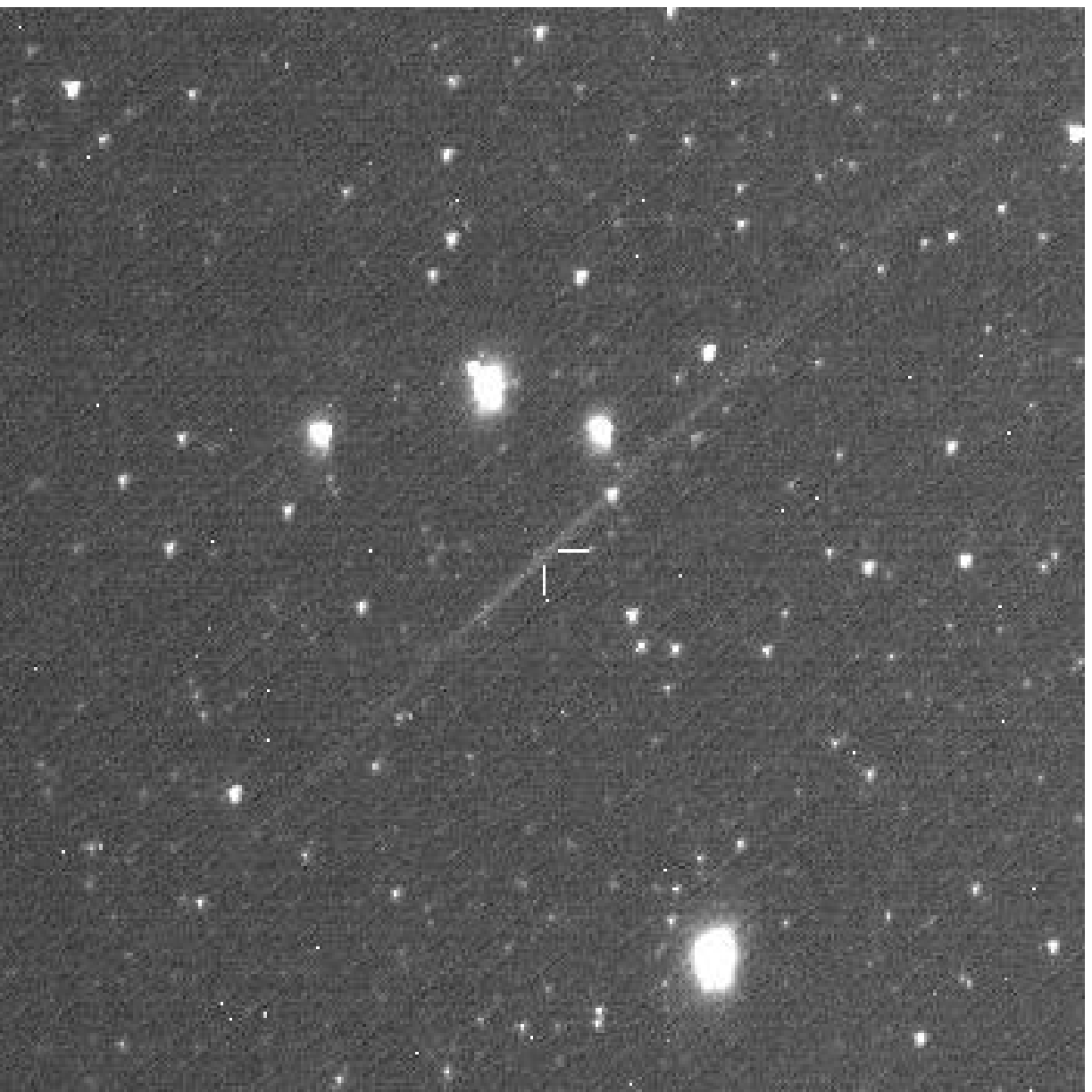,width=79mm}
\psfig{figure=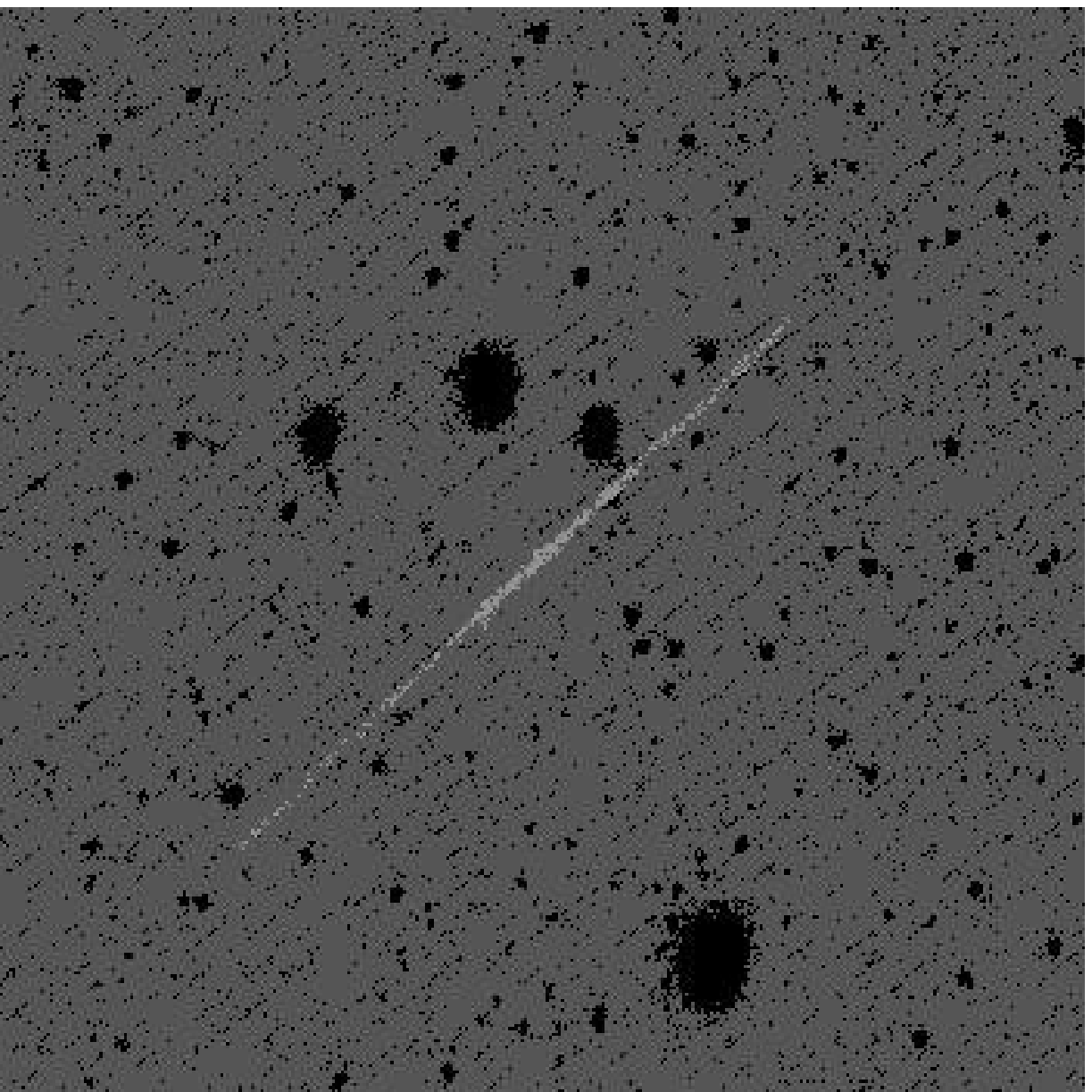,width=79mm}
\psfig{figure=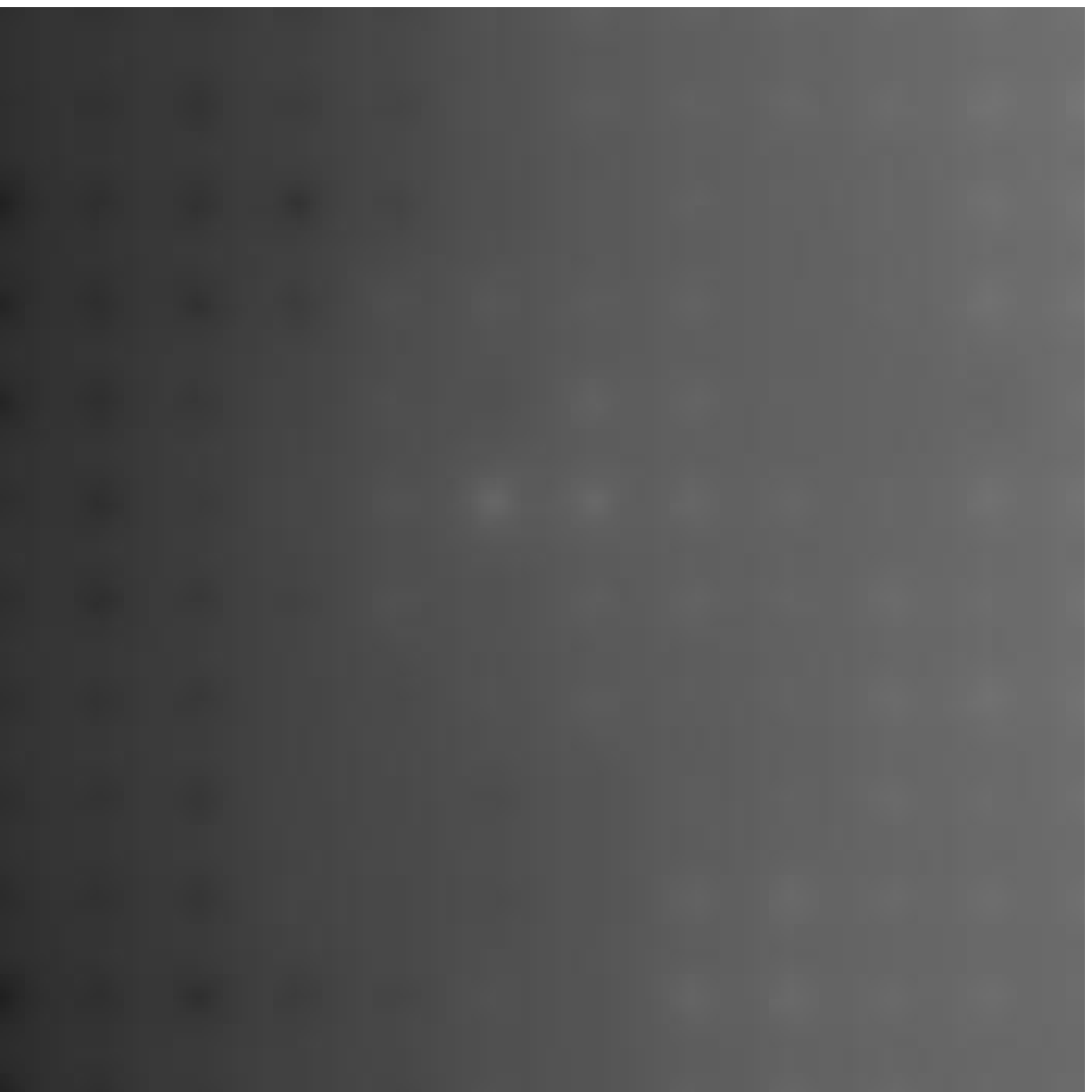,width=79mm}
\psfig{figure=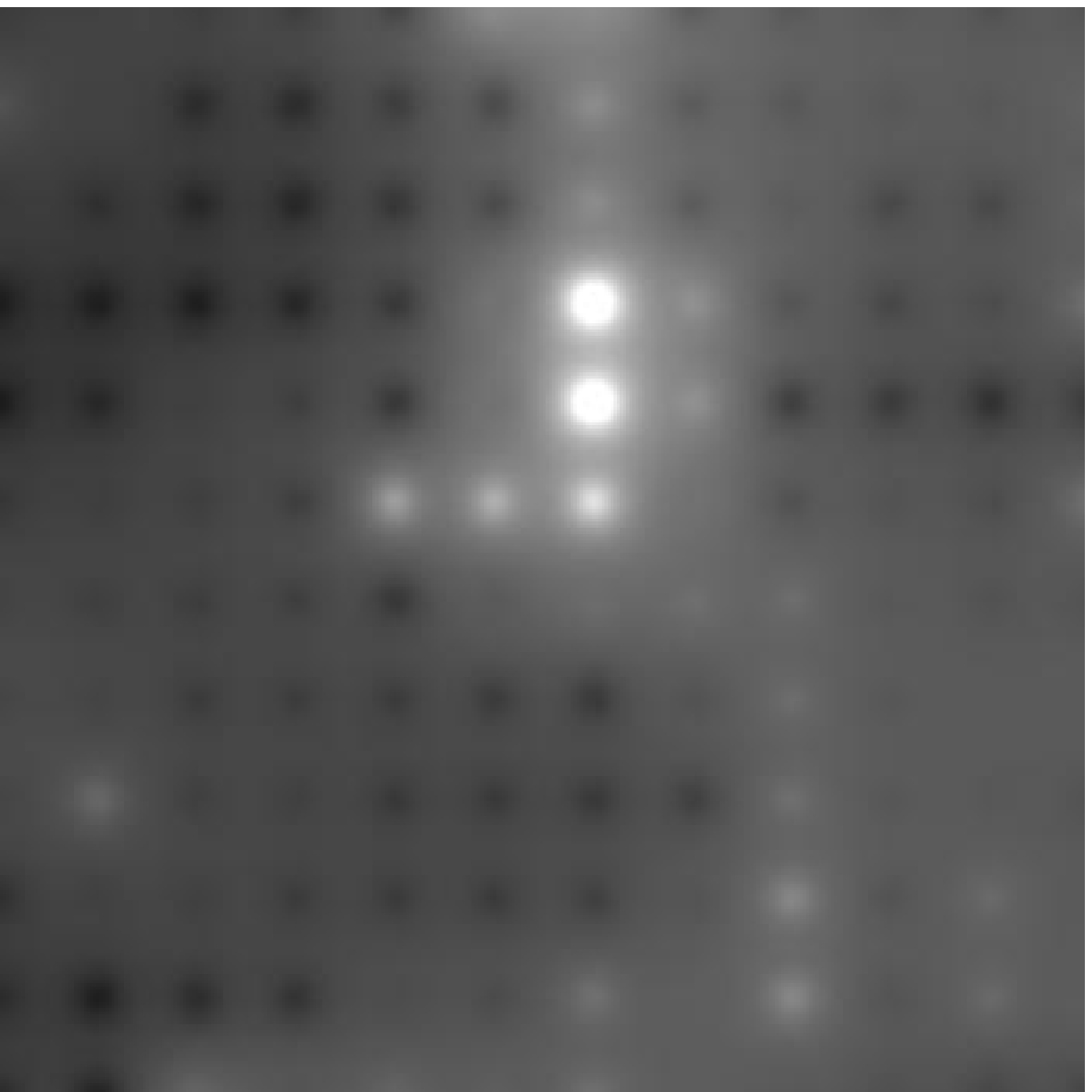,width=79mm}
\caption{  
(1) Image of a meteor on a $3 ^\circ \times 3 ^\circ $ frame ($F (i, j) $). \newline
(2) Extraction map. Black shade shows the points with $F (i, j)> n_\sigma \cdot \Sigma (i,
j) $; white shade indicates the points included into the object under study. Grey shade indicates
the background.\newline
(3) Background map - ($Bg (i, j) $) (bottom left) \newline
(4) Map of standard deviation ($ \Sigma (i, j) $) (bottom right)
}
\label{met_pro}
\end{figure}

 \paragraph {Determination of background.} To determine the background, we subdivide the
 working area into subareas of the given size. For each subarea the median and 
standard deviation are computed twice: the first time over almost all pixels 
in the given subarea, and the second time more precisely, after discarding the pixels that
deviate strongly (by more than 6-10$ \sigma $) from the preliminary background estimate.
We number all these subareas --- there are a total of
$N _ {zone} $ of them (currently $N _ {zone} =256$)--- and compute the background and standard
deviation in each subarea as 

	\begin{equation}
	\label{fon}
	\begin{aligned}
	Bg(i,j) &= \sum\limits_{k=1}^{k<N_{zone}} \frac{1}{r_k^2} \cdot Med(k)
\\
	\Sigma(i,j) &= \sum\limits_{k=1}^{k<N_{zone}} \frac{1}{r_k^2} \cdot
\Sigma(k)
	\end{aligned}
	\end{equation}

where Bg (i,j) is the background at the point with coordinates (i,j); Med (k) is the median
in k-th area used to compute the background; $ \Sigma $ is the standard deviation within the
corresponding subaera, and $r_k $ is the distance between the point (i.j) and the centre of k-th area. 
Figures~\ref {met_pro}:3-4 show the corresponding background maps.

\begin{figure}[t]
\psfig{figure=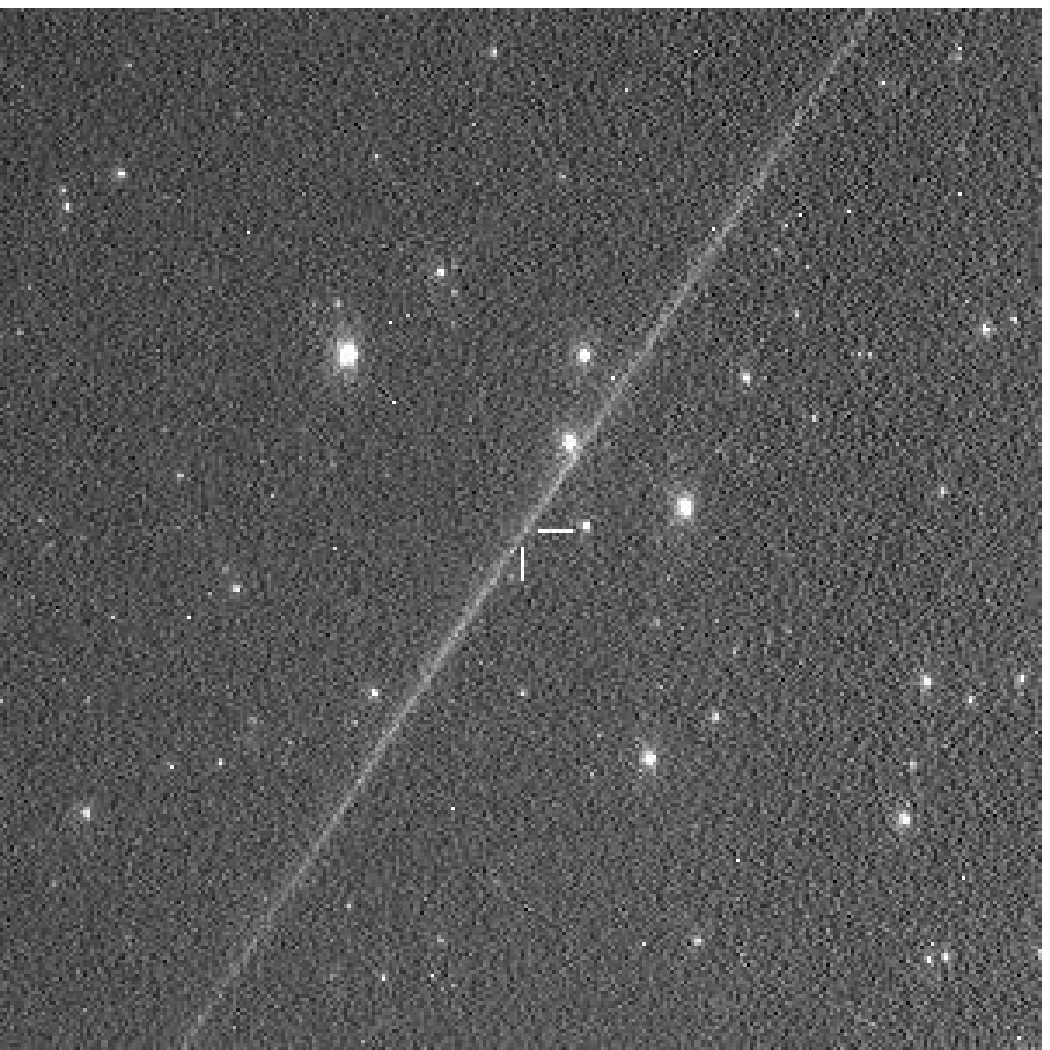,width=80mm}
\psfig{figure=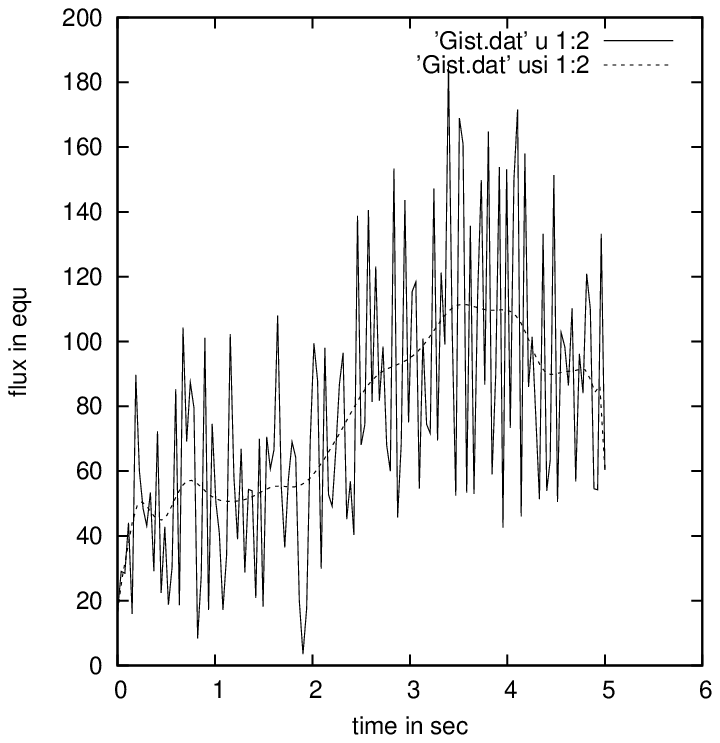,width=80mm}
\caption{Meteor and its automatically determined brightness profile.}
\label{profil}
\end{figure}

\paragraph {Extraction of strips.}

To extract a strip, we first compute the map of bright pixels exceeding the background level
by $n_\sigma \cdot \Sigma (i, j) $ or more. The optimum threshold has been shown empirically 
to be $n_\sigma=1.5$. The corresponding points are shown by black shade on the map (see fig ~\ref {met_pro}:2).

A recursive algorithm of the search for connected area is then used starting
at the centre found by SExtractor. Once the connected area is found, the main 
parameters of the object (the position of the centre, a, b, and the angle $ \theta $ between the axis of the
object and the X axis) are determined. The standard deviation $ \sigma _ {line} $ from the straight line 
drawn at angle $ \theta $ through the centre of the object  is computed if the  object found can be classified 
as a strip ($a/b> 3$). The recursive algorithm of searching for connected area is then started again,
this time within  2-3$\sigma _ {line} $ of the straight line just mentioned. The algorithm is 
allowed to jump over one to two faint pixels within  2$\sigma _ {line} $ of the straight line. 
Thus the second pass yields a completely extracted strip including faint edges in the case of meteors. 

The program then attempts to find objects located on the extension of the strip line on the subsequent
and previous frames. If such object can be found, the strip is classified as a satellite, otherwise,  as a meteor. 
The brightness profiles along the strip are then computed both for satellites and meteors (see fig. \ref {profil} ). 
Portions of images are kept for all transients.

\section {Synchronous observations of prompt grb emission.}

In this chapter we discuss synchronous grb observations. Note
that during two years of the operation of the single-channel wide-field Kislovodsk
camera we have made over 20 faster than 1 minute alert pointings published in the corresponding
GCN-telegrams GRB 070224 (gcn 6139), GRB 070223 (gcn 6131), GRB 070219 (gcn
6113), GRB 061213 (gcn 5915), GRB 061002 (gcn 5677), GRB 060929 (gcn 5657), and
others\cite{mastergrbs}.

\subsection {The importance of synchronous observations.}

As is well known, optical grb glow can be roughly subdivided into two parts: prompt
emission and afterglow. We do not discuss the afterglow phenomenon here, because it has already 
been thoroughly studied and it is not the target of wide field observations. 
Let us now consider prompt emission.

Prompt optical grb emission is the emission that appears simultaneously with gamma rays
detected by various space-based gamma observatories (Hete2, Swift, Fermi).
Unlike afterglow, it bears information about the object that gives
birth to grb, and not about its environment. There are two ways to detect prompt optical
grb emission: synchronous observations and very quick alert pointing.  To date,
about dozen  \cite{panat}  optical prompt grb have been made and most of them were 
alert based. That is to say, when prompted by a quick trigger from a gamma-ray observatory,
a robotic optical telescope  points at the object within several tens of seconds after
the corresponding gamma-ray event is detected. If the grb is long enough, its prompt emission is still
detectable by this time. Another,  less common type of observations, are synchronous 
wide-field camera observations. Perhaps the most amazing example of
the latter is the prompt GRB 080319B  detection ~\cite{racusin}. Synchronous
observations are much more efficient than alert observations for at least two
reasons.  First,  alert observations are subject to selection effects.
The point is that an alert may result in the detection of prompt emission only if the grb
lasts longer than 30-40 seconds, because of the delay due to the time it takes to perform
data processing onboard the space telescope, transmit the signal to the ground, and point the optical
telescope. Hence alerts cannot be used to detect prompt optical
emission from short grbs. It is significant that this kind of emission has never been observed and
moreover, it has never had its upper limit measured.
    
\begin{figure}[t]

\psfig{figure=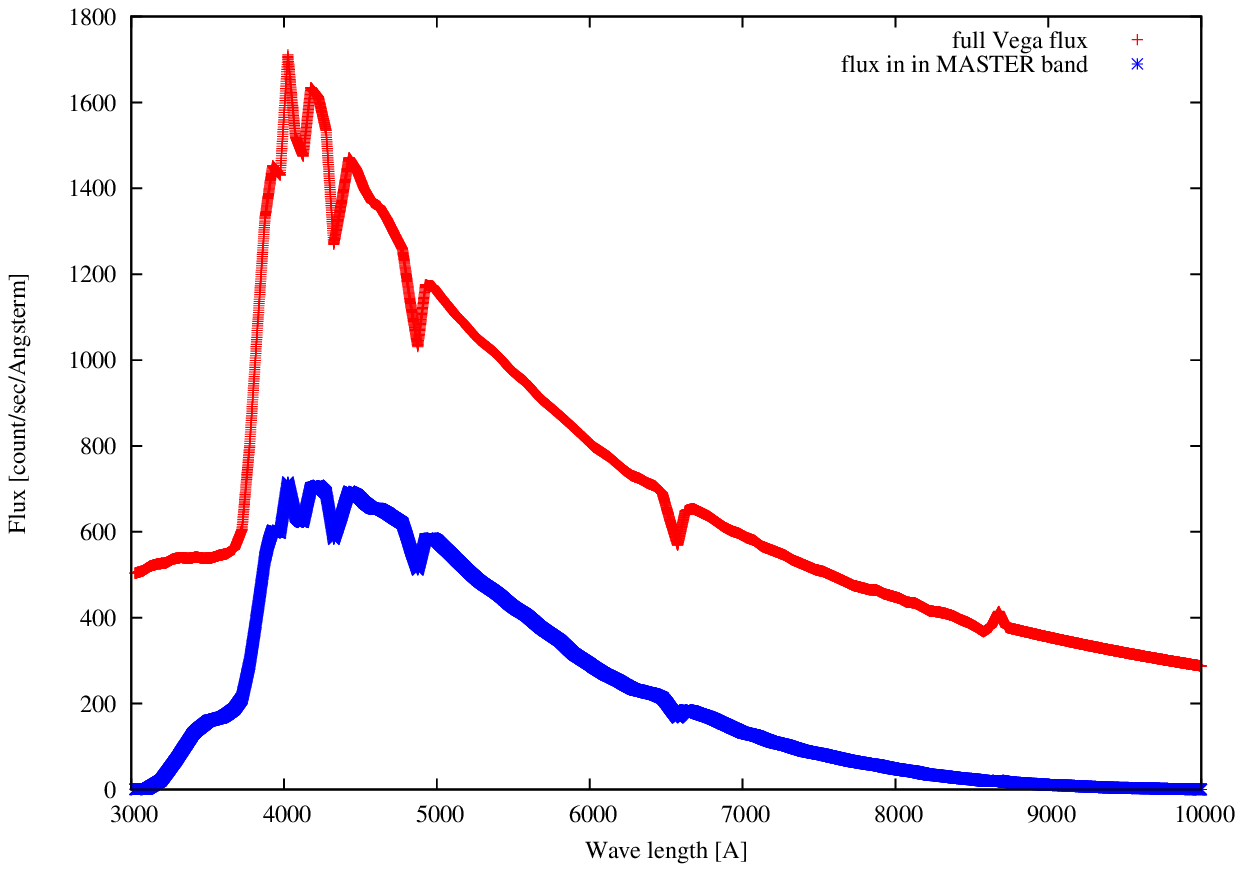,width=80mm}
\psfig{figure=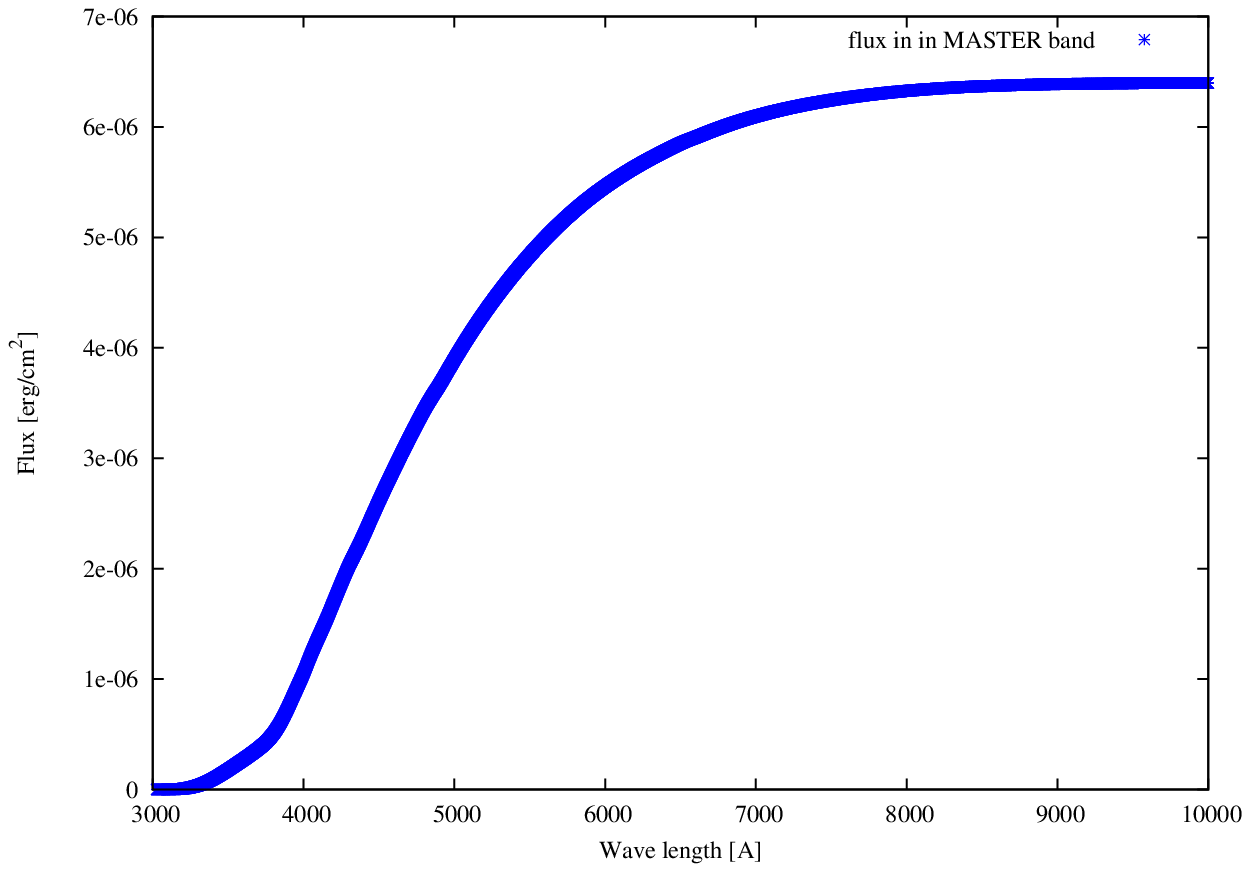,width=80mm}
\caption{The spectrum of Vega and the same spectrum as observed after passing through the MASTER VWF band  
(left panel). The cumulative curve of this spectrum in power units (the right panel).}
\label{vega}  
\end{figure}

\subsection {Energy calibrations of the MASTER-VWF4 photometric passband.}

To make any conclusions about the brightness of objects not just in terms of relative magnitudes, but also 
in terms of energy units, we must calibrate the MASTER-VWF4 zero-point. Let us determine 
the energy flux (in the units if $ erg/cm^2/s $) produced by a zero-magnitude star (e.g., Vega). 
The MASTER-VWF4 band is close to the  V band of the  Tycho2 catalogue. Figure~\ref {vega} shows the spectrum
of Vega and the same spectrum folded with  the lens passband and sensitivity curve of the
CCD. Integration over the entire spectrum  finally yields the flux produced within the
MASTER-VW4 band by a zero-magnitude star
\begin {equation}
F ^ {Vega} = \int\limits_0 ^ {\infty} f_v (\nu) \cdot \nu \cdot d\nu = 6.4 \pm
0.1 \cdot 10 ^ {-6} erg/cm^2/s
\end {equation}

\subsection {Synchronous GRB observations with MASTER VWF very wide-field  cameras}.

MASTER VWF cameras perform nightly monitoring of the sky with the aim to detect optical transients. 
These cameras may operate in two different modes.  In the first mode, which is meant for
searching for optical transients, two cameras are pointed to the same sky area in order to
implement the coincidence scheme (see fig. ~\ref
{met_par}, ~\ref {sat_par}). In the second mode the fields of view of the cameras are set apart
as far as possible in order to provide a combined field of view of more than 4000 and 2000 deg$^2$ 
for the Kislovodsk and Irkutsk observatories, respectively. The second mode is dedicated  for synchronous
observations of gamma-ray bursts.

\subsubsection {The GRB081102 and GRB090424 gamma-ray bursts recorded by Swift spacecraft}

GRBs recorded by the Swift gamma-ray observatory satellite have very accurate coordinates,
allowing the location of the burst origin on very wide-field cameras to be determined
with the highest precision possible (i.e., to within  less than one-pixel large error box).

\begin{figure}[t]
\psfig{figure=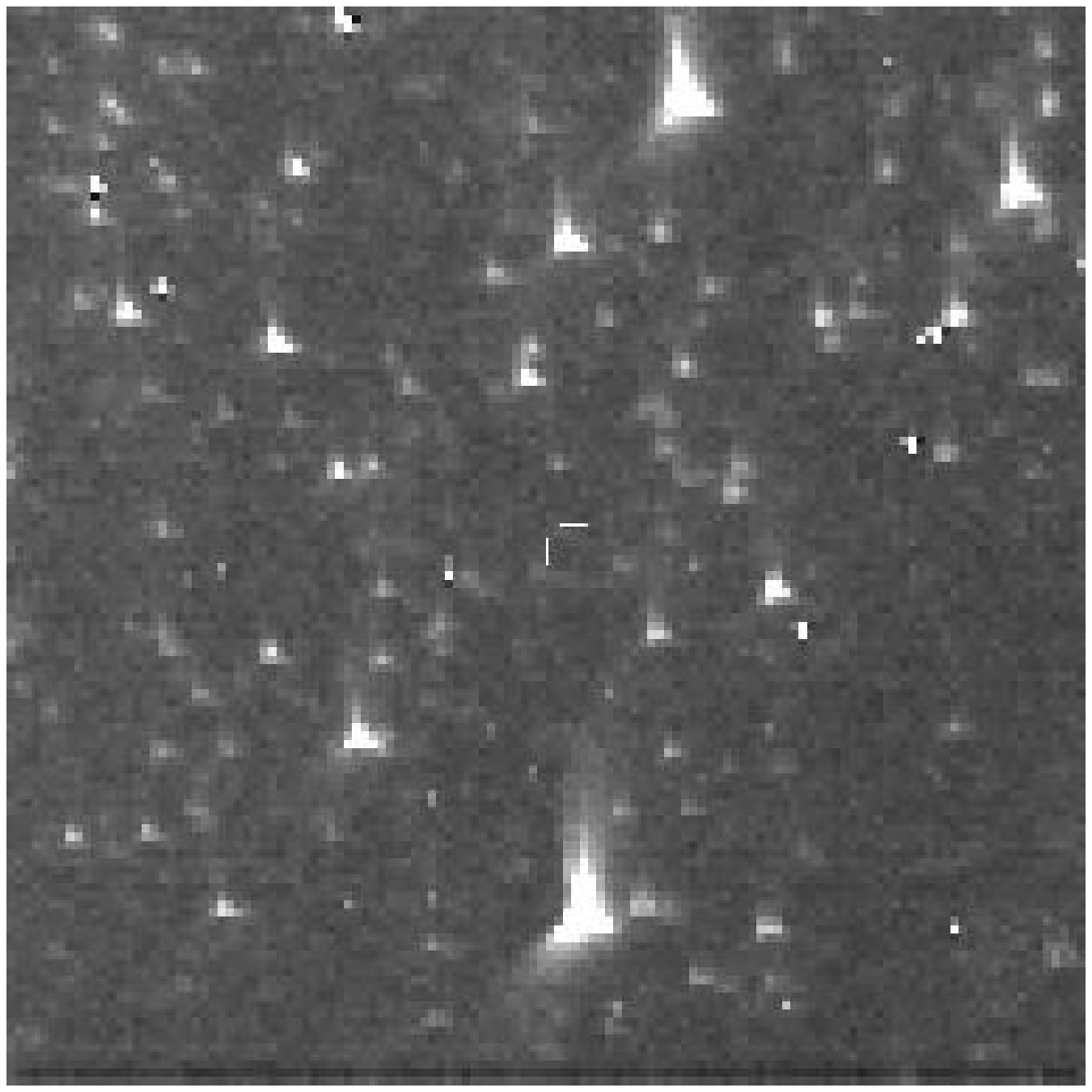,width=52mm}
\psfig{figure=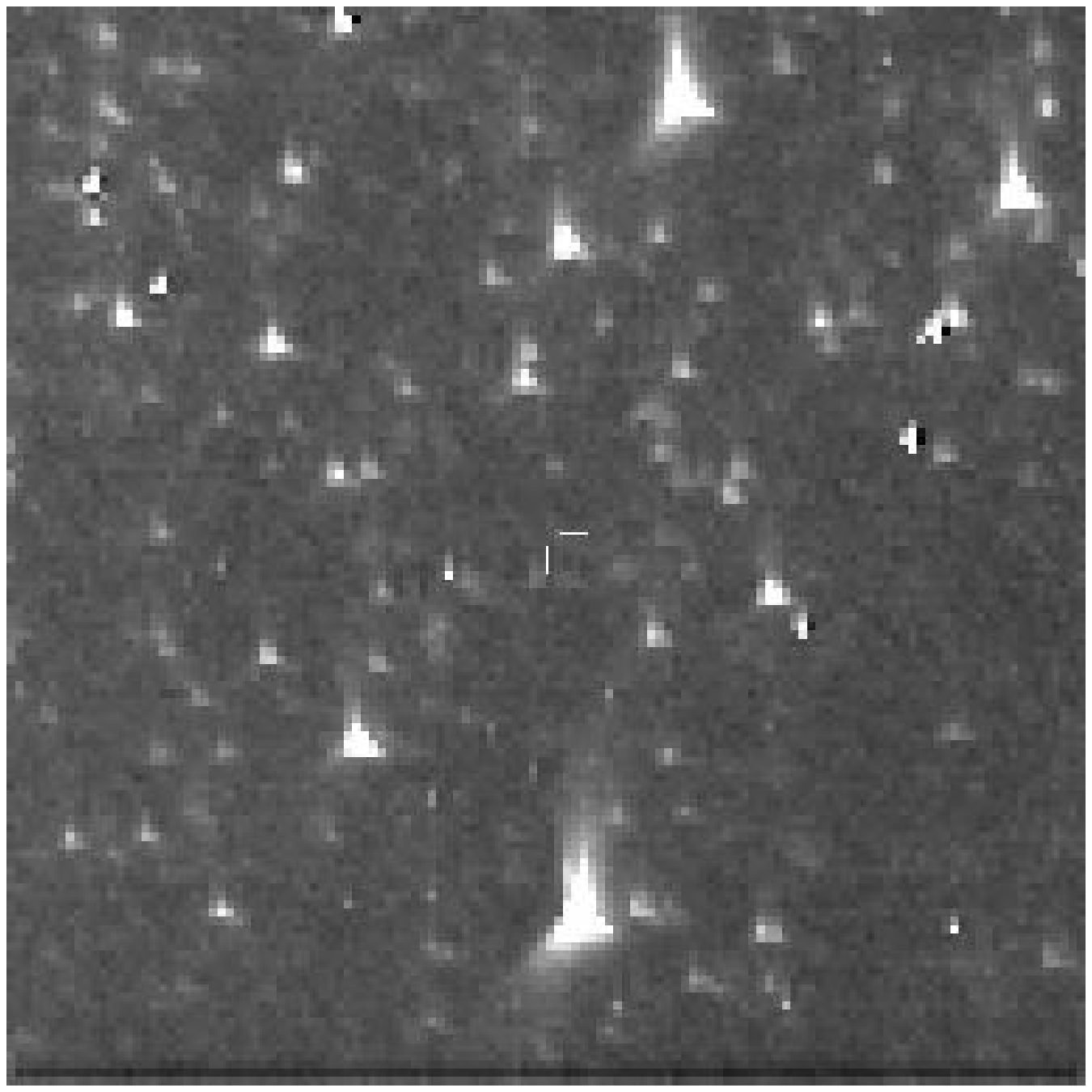,width=52mm}
\psfig{figure=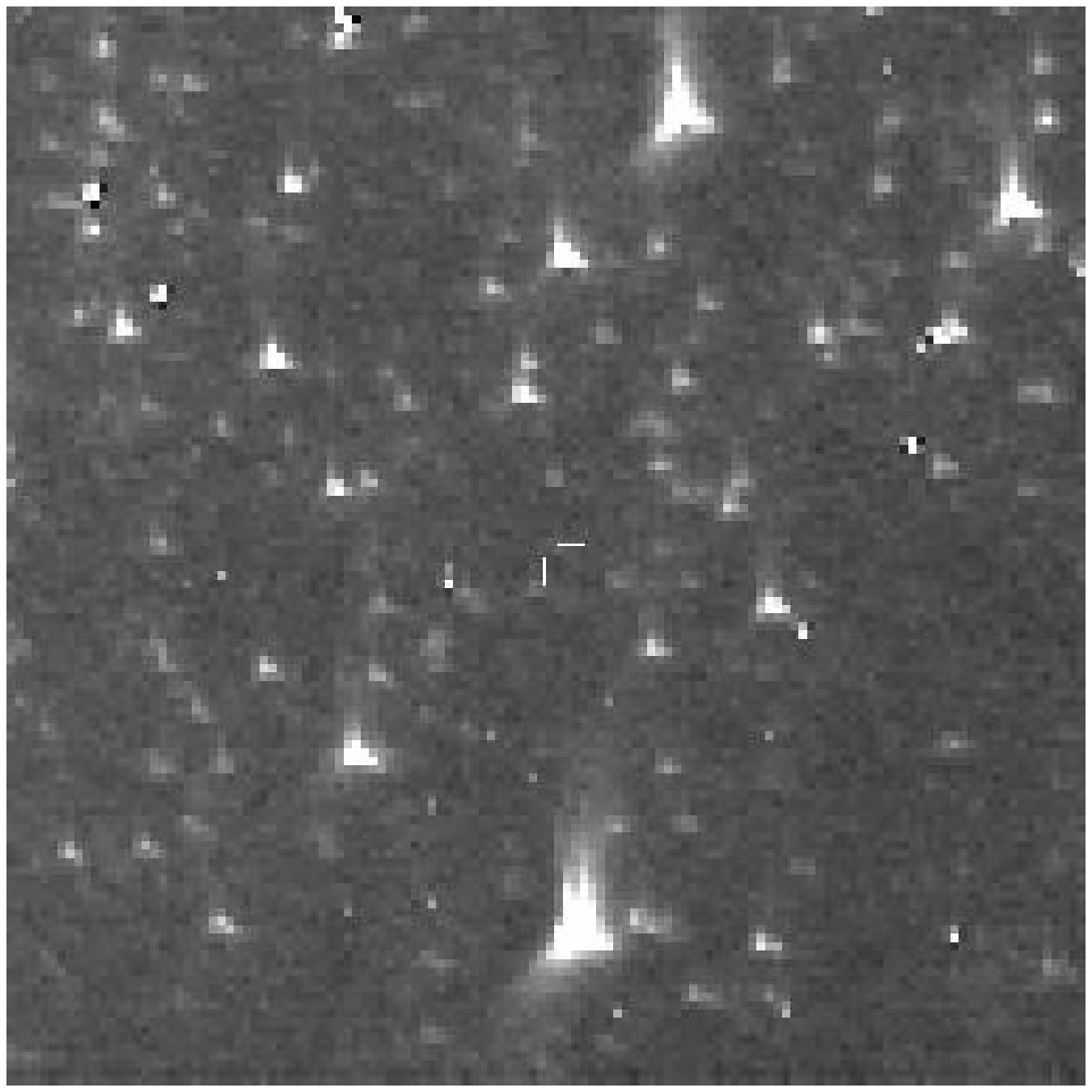,width=52mm}
\caption{Coadded images taken  30 seconds before (left), during (centre), and
30 seconds after (right) GRB081102. Each frame is a coadd of 12 images (six images  from
each camera) -- 60 seconds exposure for each frame. Animation is available here:
\itshape http://observ.pereplet.ru/images/GRB081102/grb\_film.html 
}
\label{grb081102}
\end{figure}

\paragraph {Long gamma-ray burst GRB081102} (see fig. ~\ref {grb081102} ) was
recorded by Swift gamma-ray space  observatory on November, 2nd, 2008 17:44:39.5 UT \cite {gcn8470}. 
Although it was a  typical long ($T _ {90} =63 sec $) gamma-ray burst, the 
telegram from Swift was issued  only after 15
minutes. This delay prevented early observations by  automated telescopes.  However,
MASTER VWF-4 in Kislovodsk covered the entire error box at the edge of its field of view.
MASTER VWF-4 operated in the survey mode  with aligned cameras, allowing  the grb error-box to be
recorded simultaneously in two channels. This area has been observed for two hours before and
seven hours after the grb with no time gaps \cite{gcn8471}.

\begin{figure}[t]
\psfig{figure=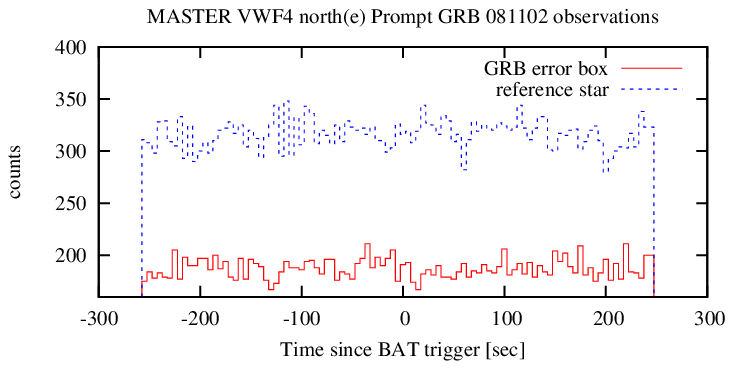,width=52mm}
\psfig{figure=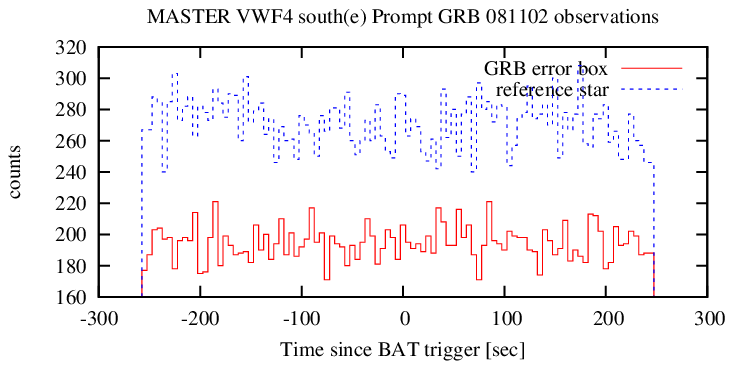,width=52mm}
\psfig{figure=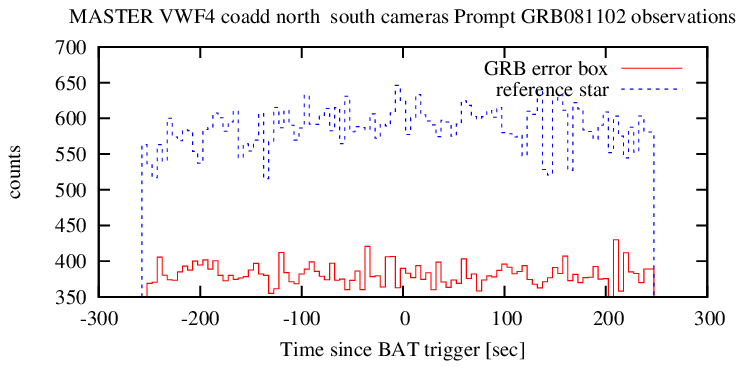,width=52mm}
\caption{GRB081102. Hash (red) from the Northern and Southern cameras and from
the two cameras simultaneously inside the GRB081102 error-box and the signal from the $V _ {ref} =11.5^m $ 
reference star (blue)
}
\label{lc}
\end{figure}

Unfortunately, no  optical counterpart could have been recorded from this burst 
(see fig. ~\ref {grb081102}, ~\ref {lc}). The lack of optical counterpart may have been due to
extremely strong absorption toward the burst (\ref{pogl}). However, a very high
upper limit (for synchronous (!) observations with very wide-field
cameras)  of $V _ {grb081102} ~ <~13.0^m $ has been obtained, which is the highest 
synchronous upper limit in history \cite{gcn8516}.

Let us now compare  GRB081102 with GRB080319B from which bright prompt
optical emission $V=5.3^m $ (\cite {racusin}) has been recorded. Consider the
$F _ {opt}/F_\gamma $ ratio (where $F _ {opt}$ and $F_\gamma $ are the optical and gamma-ray flux,
respectively) for both bursts.  The gamma-ray spectrum is known for all bursts discussed in this
paper and therefore we reduce  $F_\gamma $ to the flux in the $15-150 keV $ energy interval 
for the sake of uniformity.

Let us estimate $F _ {opt} ^ {grb081102} $. The V-band extinction can be estimated by the following
simple empirical formula \cite
{zaspos}

\begin{equation}
\label{eqpog}
\tau_V = 5.2\cdot10^{-22} N_H
\end{equation}

Given the known hydrogen column density toward GRB081102, $N_H=4.9*10 ^ {21} cm ^ {-2} $ \cite {calb},
we can use the inferred optical depth  and Pogson's formula to estimate 
$F_\gamma^{grb081102}=2.3 \cdot 10^{-6} erg/cm^2$
\cite{gcn8468}   or

\begin{equation}
\label{pogl}
\tau_V^{grb081102} = 5.2\cdot10^{-22} \cdot 4.9\cdot10^{21} = 2.548 => \delta m
=  2.8^m
\end{equation}

\begin{equation}
\label{rel_grb08110}
\frac {F_{opt}^{grb081102}}{F_\gamma^{grb081102}} < \frac
{2.512^{m^{Vega}-(m^{grb081102}-\delta m)} \cdot F^{Vega}\cdot
T_{90}^{grb081102}} {F_\gamma^{grb081102}}
\end{equation}
$$
\frac {F_{opt}^{grb081102}}{F_\gamma^{grb081102}}
<\frac{2.512^{0-13.0+2.8}\cdot6.4\cdot10^{-6} erg/s/cm^2 \cdot 63s}{2.3 \cdot
10^{-6} erg/cm^2} = \frac {1}{83}
$$

The "less than" sign is used because only the upper limit could have been determined for grb081102.
For  GRB080319B the column density is $N_H=9.2*10 ^ {20} cm ^ {-2} $ \cite {racusin}, i.e.,
five times lower than toward GRB081102 $ \tau_V ^ {grb080319B} =0.48$ => $
\delta m = 0.5^m $. Hence

\begin{equation}
\label{rel_grb080319B}
\frac {F_{opt}^{grb080319B}}{F_\gamma^{grb080319B}} = \frac
{2.512^{m^{Vega}-(m^{grb080319B}-\delta m)} \cdot F^{Vega}\cdot
T_{90}^{grb080319B}} {F_\gamma^{grb080319B}}
\end{equation}
$$
\frac {F_{opt}^{grb080319B}}{F_\gamma^{grb080319B}} =
\frac{2.512^{0-13.0+2.8}\cdot6.4\cdot10^{-6} erg/s/cm^2 \cdot 50s}{8.25 \cdot
10^{-5} erg/cm^2} = \frac {1}{22}
$$

It thus follows that the fraction of prompt optical emission for GRB081102 is at least four times smaller
than for grb080319B.

\begin{figure}[t]
\psfig{figure=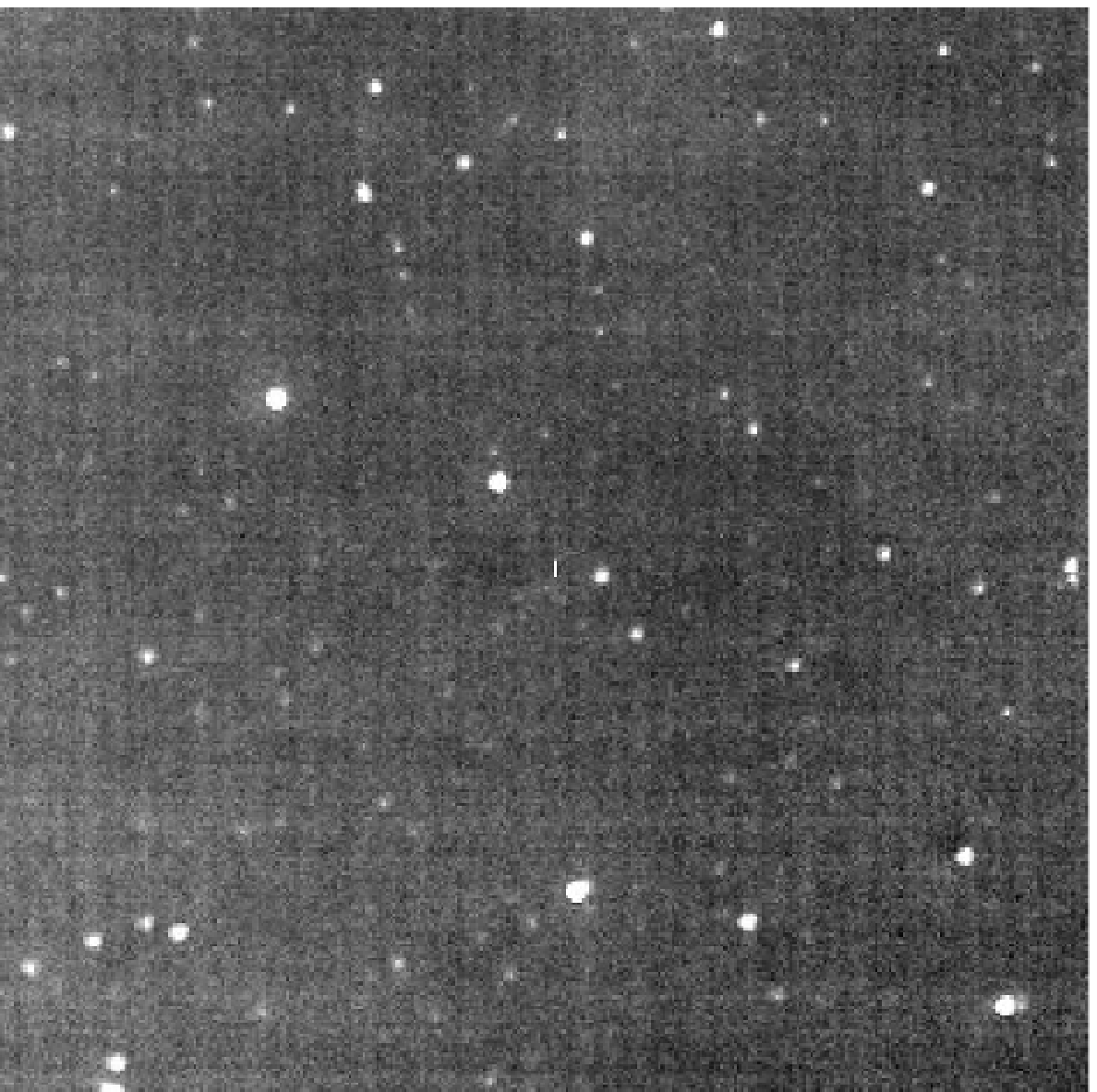,width=52mm}
\psfig{figure=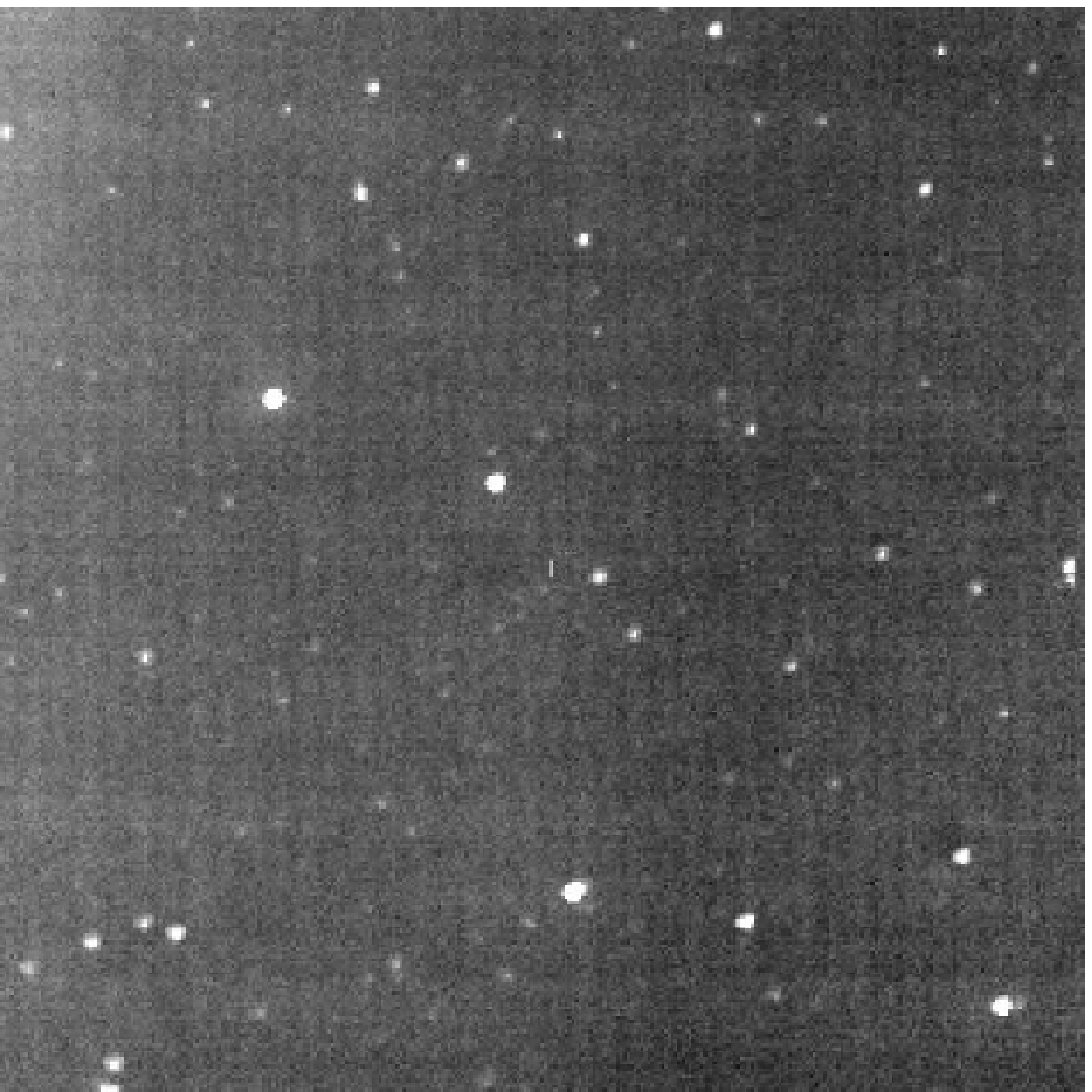,width=52mm}
\psfig{figure=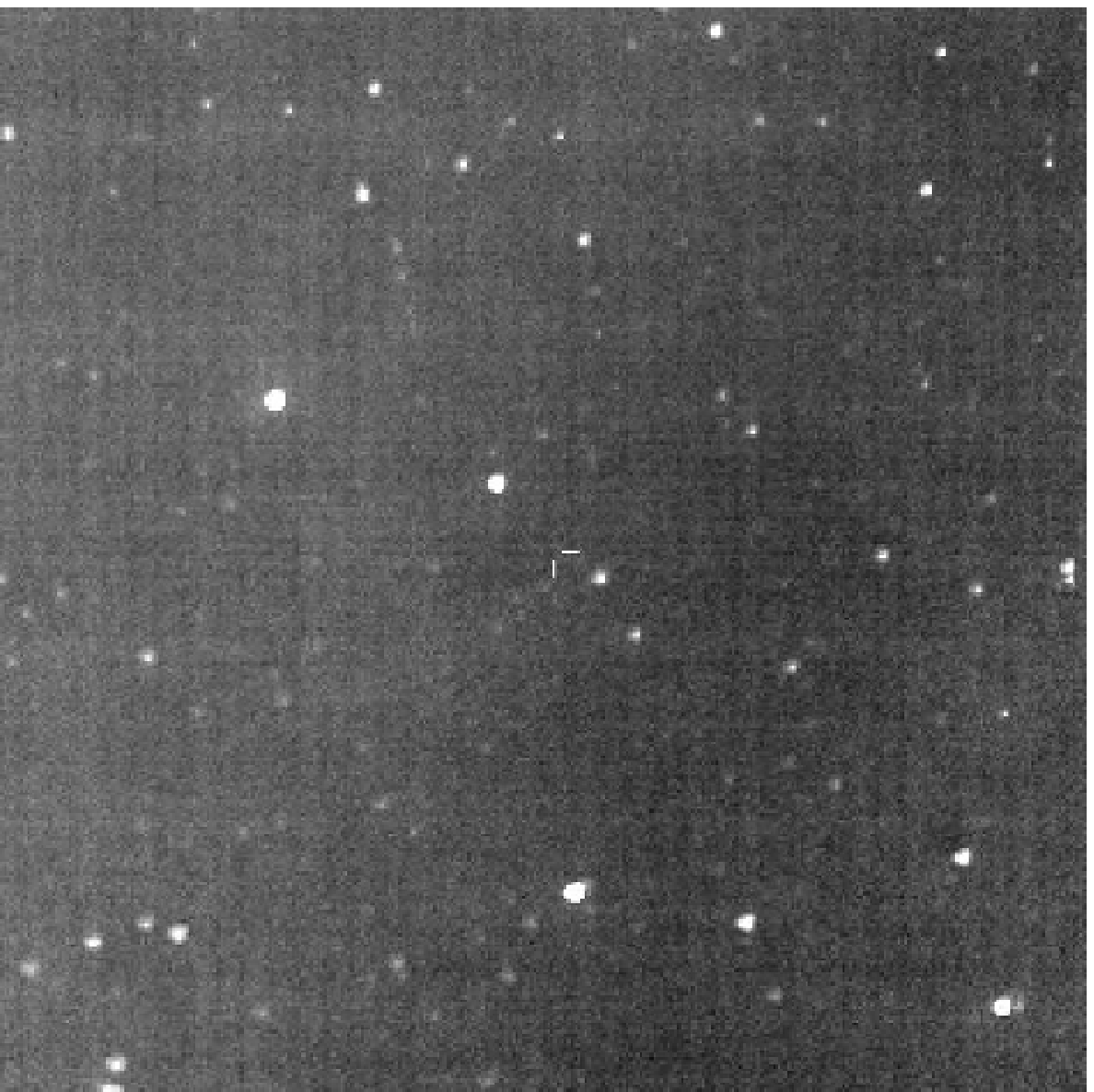,width=52mm}
\caption{A 6x6 deg$^2$ field around the GRB090424  Swift XRT position. One-minute exposure images taken
before, during and after $T_{GRB}$. Clouds can be seen  on the images, resulting in the optical
limit of $V<9.7^m$ on each frame. Animation available here: \itshape
http://observ.pereplet.ru/images/GRB090424/grb\_film.html}

\label{grb090424}
\end{figure}

\paragraph {The long gamma-ray burst GRB090424} (fig ~\ref {grb090424}) was recorded by
Swift observatory on April, 24th, 2009 14:12:09.33 UT \cite {gcn9231}. At that time MASTER
VWF-2 in Irkutsk operated in the alert mode  (with  cameras pointed apart)
and GRB090424 happened to be at the centre of the field of view of one of the cameras. It was also
a typical long ($T _ {90} =48 sec $) gamma-ray burst and  UVOT detected afterglow
with unfiltered magnitude of about $15.3^m $ 167 seconds after the burst \cite {gcn9234}.
MASTER VWF-2 observed the grb error-box without time gaps for 1.2 hours before, during
 and 1.5 hours after the gamma-ray burst with 1-second exposures. Unfortunately, 
weather conditions were far from favourable, resulting in a not so high upper limit of $V _ {1
sec} <8.0^m $ and $V _ {60 sec} <60.0^m $ for a coadd of 60 images. \cite
{gcn9252} and \cite {gcn9233}. According to both Swift/BAT \cite {gcn9231} data and 
FERMI observatory data converted to the $15-150keV $ energy interval \cite
{gcn9230},   $F_\gamma ^ {grb090424} =2.1e-5$, implying that

\begin{equation}
\label{rel_grb090424}
\frac {F_{opt}^{grb090424}}{F_\gamma^{grb090424}} < \frac {1}{530}
\end{equation}

Note that despite the rather modest  upper limit the burst was very powerful and even such
an upper limit restricts substantially the possible flux.

\begin{figure}[!t]
\psfig{figure=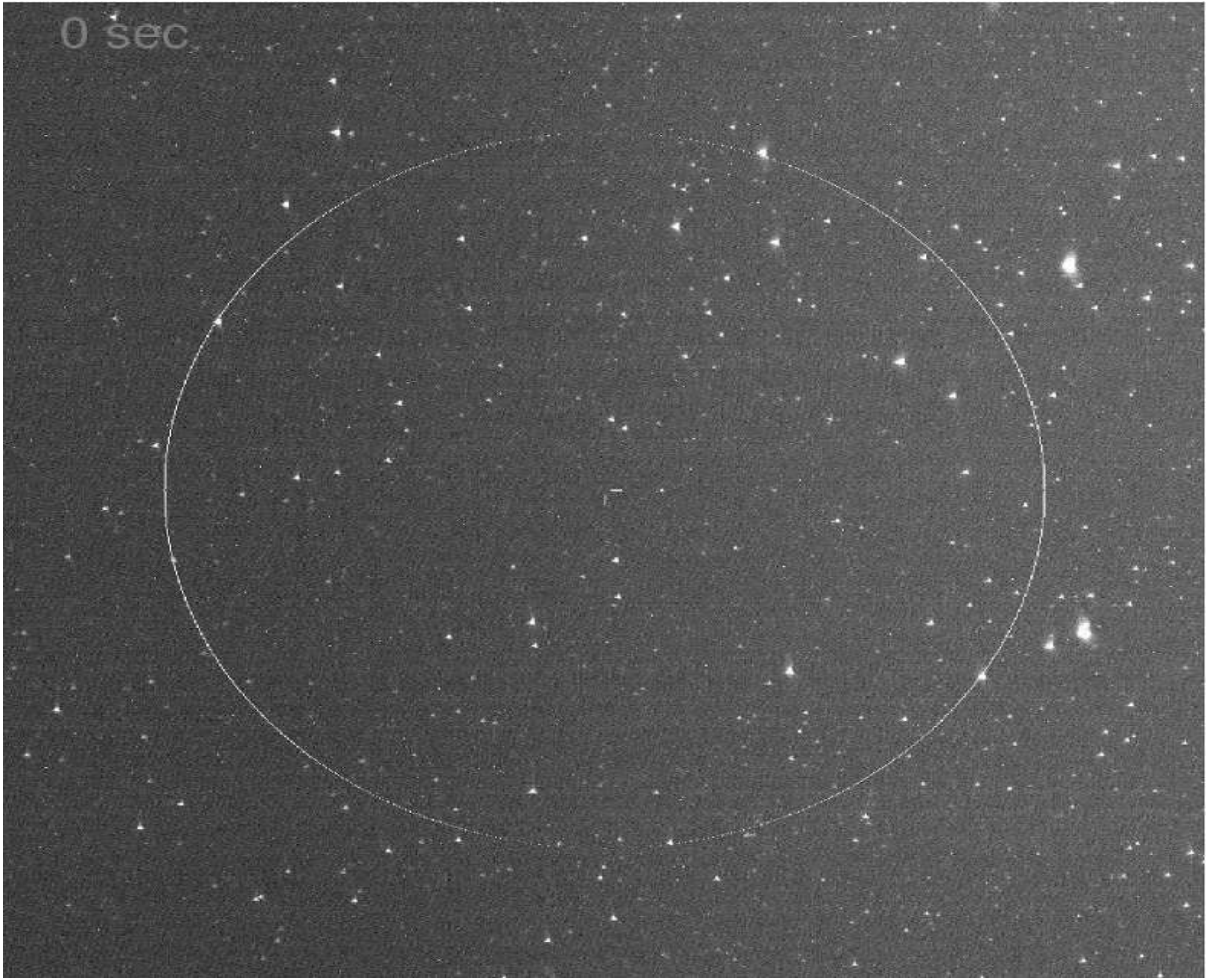,width=160mm,height=130mm}
\caption{The most recent error box of GRB081130B $\alpha=00^h 56^m 20^s$
$\delta=+04^o12'$, error-box radius is $R=3.5$ degrees \cite{8593}. The centre
of the field of view of the camera that which recorded the burst pointed at  $\alpha=01^h 01^m 56^s.49$ 
$\delta=+19^o 20' 38''.88$. The image size is $10^o \times 10^o$. Animation
available here: \itshape
http://observ.pereplet.ru/images/GRB081130B/grb\_film.html}
\label{grb081130}
\end{figure}

\subsubsection{Gamma-ray bursts recorded by FERMI observatory}

Unlike Swift, the FERMI observatory does not specialise in gamma-ray burst
observations. Its onboard equipment allows the burst coordinates to be determined only to within
several degrees or even worse. FERMI currently issues two to three times  more alerts than 
Swift and is capable of recording harder (up to 10 MeV) radiation (Swift has an upper sensitivity
limit of  about  300 keV), and responds to short hard gamma-ray bursts,  whose prompt emission has
so far never been detected. Given the large error of the coordinate measurement, gamma-ray bursts detected
by FERMI observatory may at present (and, very likely, also in the future) have their prompt emission observed
only with wide-field cameras, such as MASTER-VWF.

Until now (July, 1 2009), a total of five FERMI GRB error boxes have overlapped with the  MASTER-VWF field of view,
and two more (GRB081215 and GRB090526) proved to be outside the field of view after a very substantial final 
final refinement of their coordinates.

\begin{figure}[t]
\psfig{figure=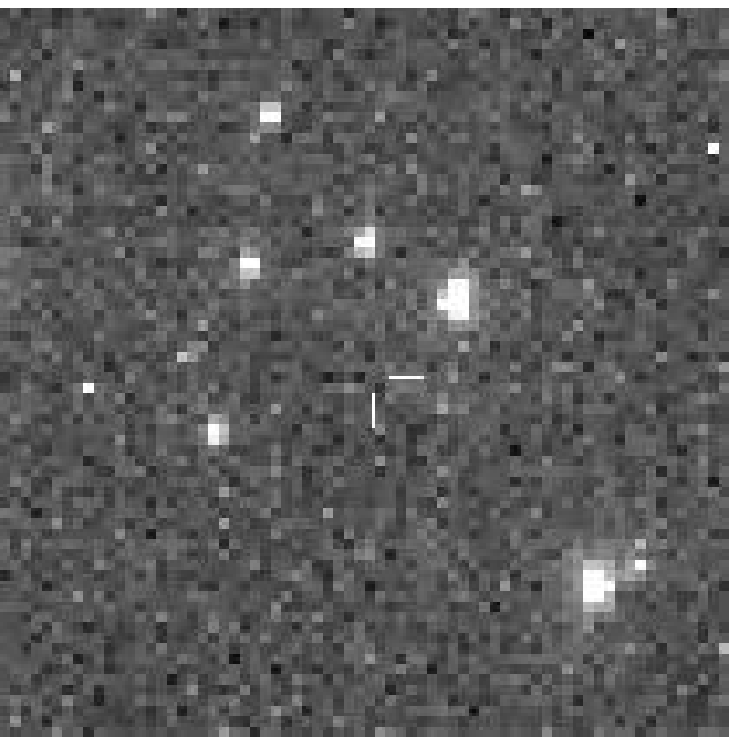,width=52mm}
\psfig{figure=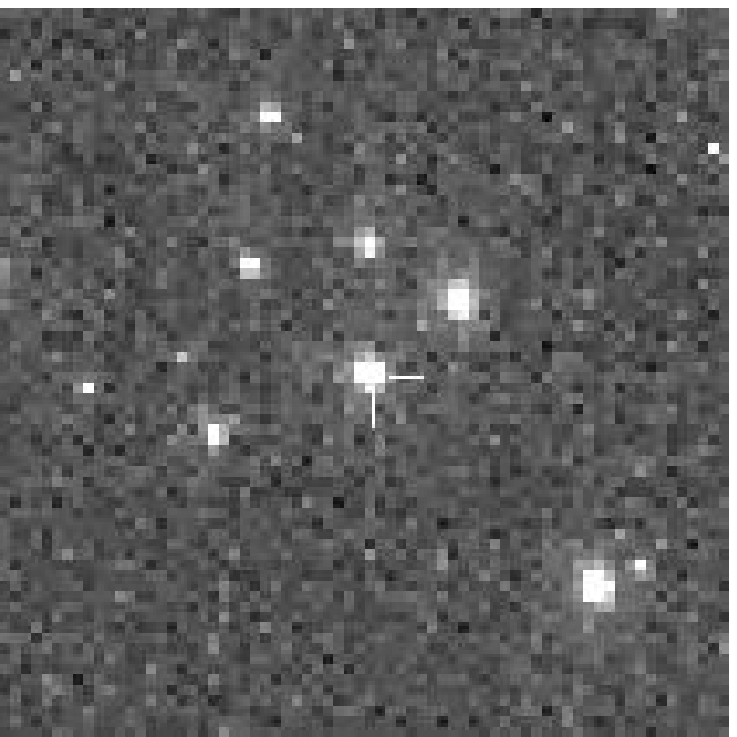,width=52mm}
\psfig{figure=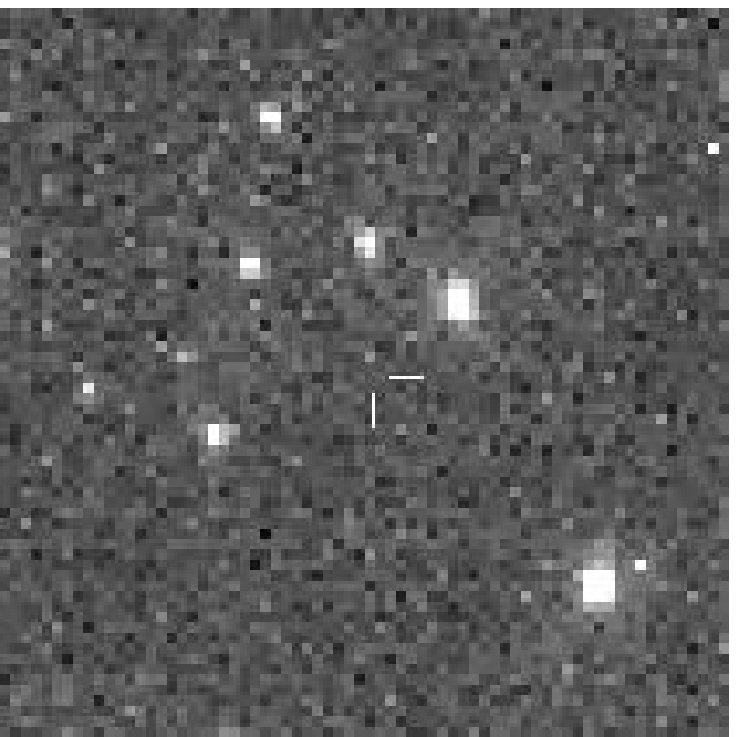,width=52mm}
\caption{
Transient object found simultaneously with GRB081130B at $\alpha=01^h10^m 20^s $ $ \delta = + 18^o44'35 "$,
14 degrees from the centre of its last GRB081130B localisation. A detailed analysis showed the object to be
a short flash produced by the "MOLNIYA" high-orbit satellite.
}
\label{tr_grb}
\end{figure}

\paragraph {GRB081130В} was the first burst recorded by FERMI and
synchronously observed by the MASTER VWF4 facility in Kislovodsk. Its coordinates and
error-box have been refined repeatedly (with a characteristic scatter of $ \pm 10$
degrees), but always fell within the field of view of our cameras. 
Figure~\ref {grb081130} shows the final localisation of this  gamma-ray
burst. The error box of grb081130 has been observed without time gaps for several hours
before, during, and for hours after the grb. 

No proper candidates could have been found inside the error box. However,
given the rather large uncertainty of the coordinates measured by FERMI,
we extended the search area for transient candidates to the entire frame. 
As a result, a very interesting object resembling a  grb optical counterpart has been
found 14 degrees from the localisation centre synchronously with grb (fig.
\ref {tr_grb}). However, after cross identification with the catalogue of artificial satellites
this transient was found to coincide with the high-orbital satellite, which
occasionally produces such flashes. Thus a rather high upper limit of $V _{grb081130B}> 12^m.0$
has been obtained again after the coadd of four images ($T _ {grb081130B} =15-20 s $ 
taken in different channels \cite{8593} ) 
\cite{gcn8585}.

When converted to the 15$-150keV$ energy interval, the gamma-ray fluence from GRB081130В is
$F_\gamma ^ {grb081130B} = 1.01 \cdot 10 ^ {-6} erg/cm^2$. We then apply the same procedures as 
in the case of GRB081102 to infer $F _ {opt} ^{grb081102} = 3.82 \cdot 10 ^ {-9} erg/cm^2$, 
implying a flux ratio of:
$
\frac {F _ {opt} ^ {grb081130B}} {F_\gamma ^ {grb081130B}} <\frac {1} {245}
$ \cite {gcn8597}

\paragraph{090305B}

\begin{figure}[!t]
\psfig{figure=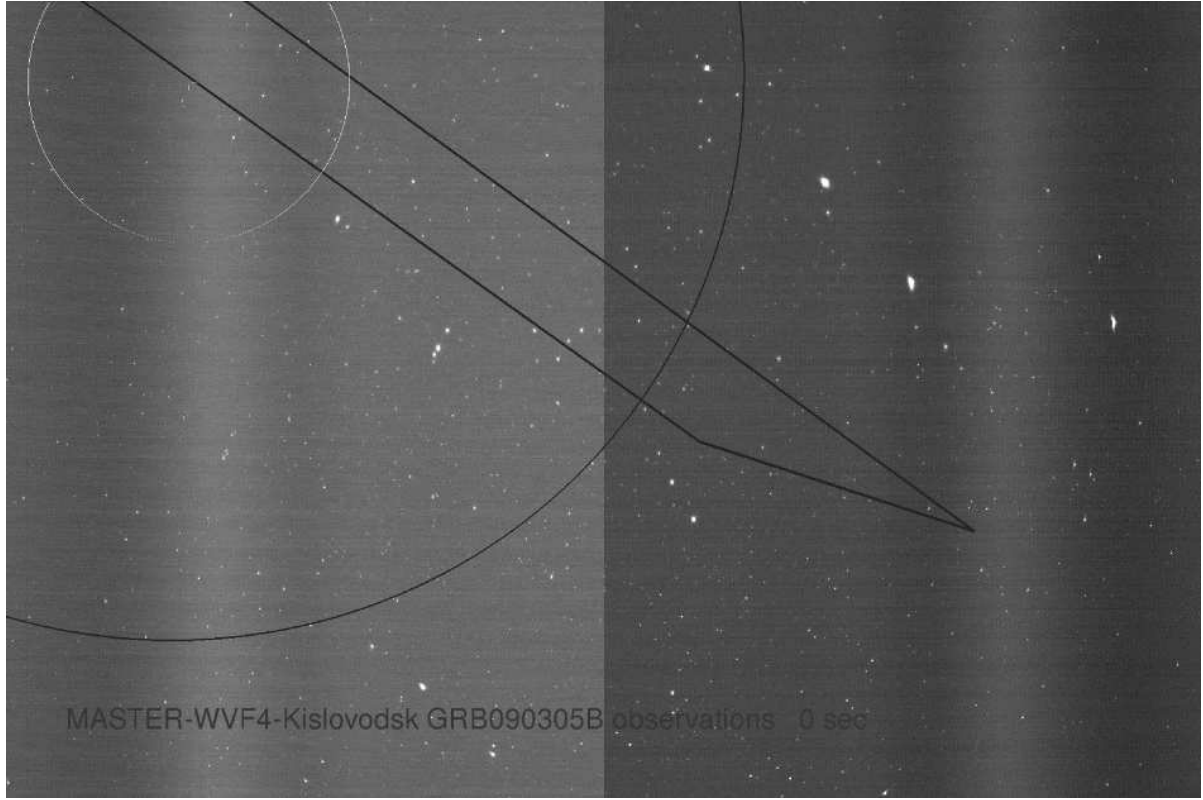,width=159mm}
\caption{MASTER-VWF4 Kislovodsk GRB090305B follow-up observation. This is a 40x24 deg$^2$ 
full-frame image. The white and black circles show the error boxes corresponding to the 1 and 3 sigma
FERMI error limits  plus the 3-degree systematic error \cite{gcn8972} ($R_{1 sigma}=5.4
deg$). The black rectangle shows the IPN triangulation error-box 
Animation available here \itshape
http://observ.pereplet.ru/images/GRB090305B/GRB090305B\_full2.gif }
\label{grb090305}
\end{figure}

The gamma-ray burst GRB090305 (Fig.~ \ref {grb090305}) was recorded by FERMI on
March, 5th, 2009 14:12:09.33 UT \cite {gcn8972}. MASTER VWF-4 in Kislovodsk
operated in the alert mode (with cameras apart) and observed 80 \% part of the
GRB090305 error-box at the edge of the field of view of one of cameras. Despite of its
rather short duration ($T _ {90} =2 sec $) \cite {gcn8972},  GRB090305 was nevertheless 
a rather long and bright gamma-ray burst. However, even the final position estimate 
has a large error-box ($ \alpha=10^h 20^m $, $ \delta=68^{\circ} 06' $ $R _ {1 sigma}
=5.4^{\circ} $) and therefore no alert observations have been made. MASTER VWF-4 has observed 
the error-box of this GRB without time gaps for 3.5 hours before, during and for two hours after
the with 1-second exposures. No optical transients close
to $T_{GRB} $ could be found and we obtain an upper-limit of $V _ {1 sec} <9.5^m $ \cite {gcn9004}.
$ \frac {F_{opt}^{grb090305B}}{F_\gamma^{grb090305B}} < \frac {1}{121} $

\begin{figure}[!t]
\psfig{figure=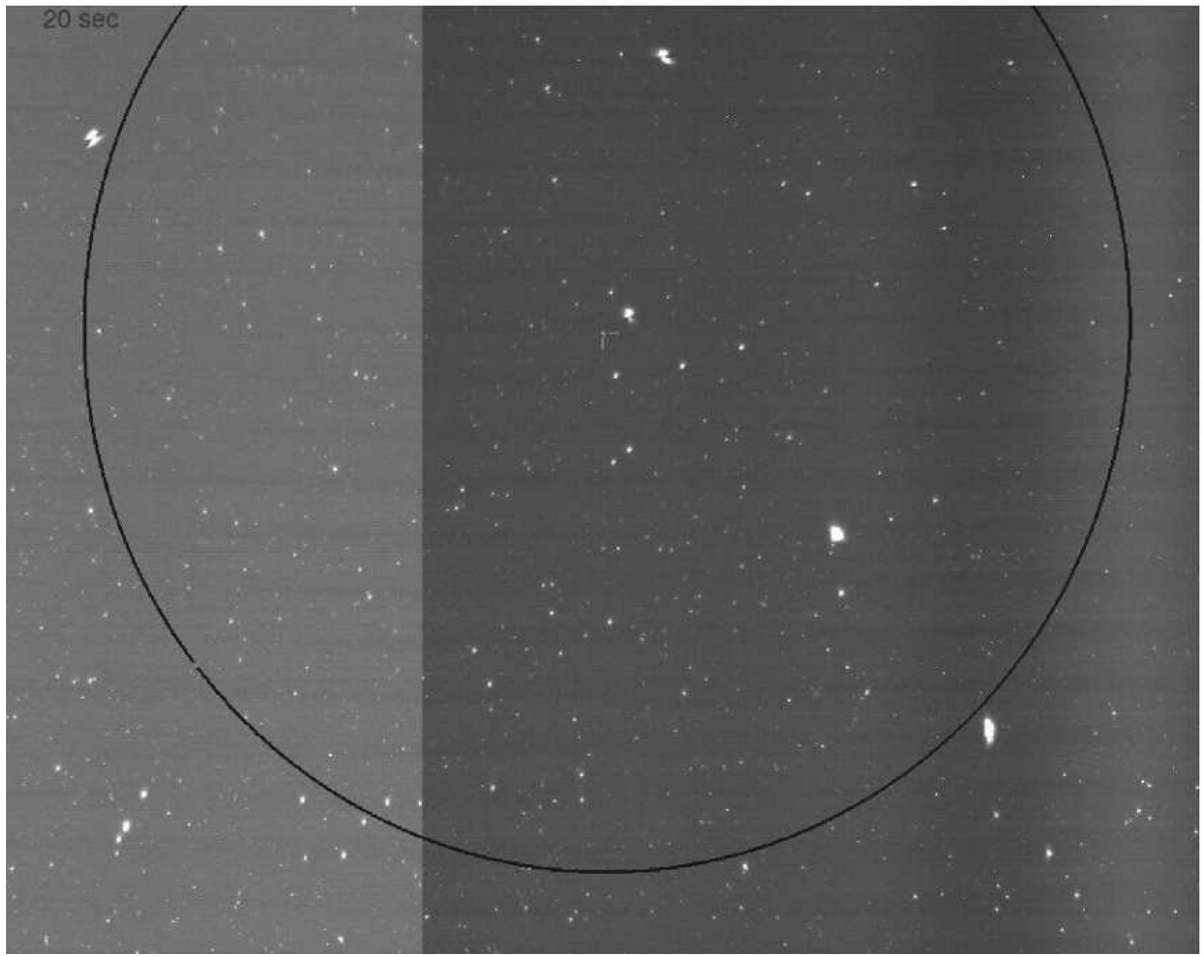,width=159mm}
\caption{
MASTER-VWF4 Kislovodsk GRB090320B follow-up observation. This is a 40x24 deg$^2$  full-frame 
image. The black circle shows the error box corresponding to the 3 sigma
FERMI error limits  plus the 3-degree systematic error  ($R=12.6 deg$)\cite{gcn9020}. 
Animation available here \itshape
http://observ.pereplet.ru/images/GRB090320B/GRB090320B\_60sec.gif 
}
\label{grb090320}
\end{figure}

\begin{figure}[!t]
\psfig{figure=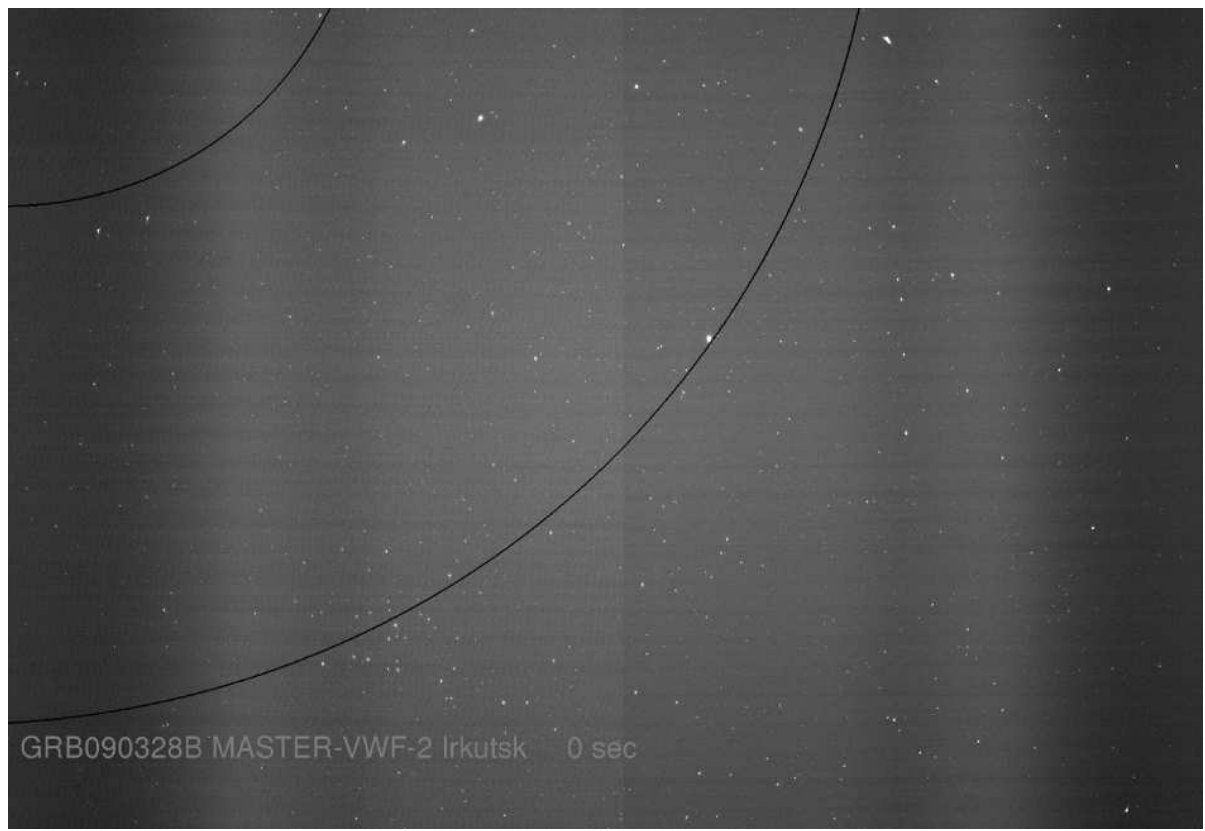,width=159mm}
\caption{MASTER-VWF4 follow-up observations of GRB090328B in Kislovodsk. This is a 40x24 deg$^2$  full-frame 
image. The black circle shows the error box corresponding to the 3 sigma
FERMI error limits  plus the 3-degree systematic error ($R=12.6 deg$)\cite{gcn9065}. 
Animation is available here \itshape
http://observ.pereplet.ru/images/GRB090320B/GRB090320B\_60sec.gif  
}
\label{grb090328B}
\end{figure}

\paragraph{GRB090320B} 
The MASTER VWF-4 system in Kislovodsk observed synchronously  80 \% of the
GRB090320B (see fig.\ref {grb090320}). This moderately bright burst has been detected  by FERMI observatory at
19:13:46.1 UT on 20 March 2009 at $ \alpha=12^h 15^m.3$, $ \delta=57^d 33 ' $ with an uncertainty of 
12.6 degrees (this includes the 3-degree systematic error) \cite {gcn9020}. 
No  optical candidates could be found for this burst, which was long enough ($T _ {90} =60 s $) 
for its  upper-limit $V <11^m $ to be estimated on 60-sec. exposure sets ~\cite{gcn9038}.
On the whole, this burst turned out to be rather faint with
$
\frac {F _ {opt} ^ {grb090320B}} {F_\gamma ^ {grb090320B}} <\frac {1} {57}
$

\paragraph{GRB090328B} 
Finally, GRB090328B is the last of the bursts considered here to deserve a separate discussion.
It had 20\%  of its error box observed by MASTER VWF-2 in Irkutsk (Fig.~\ref {grb090328B}). GRB090328B was 
short compared to all other long bursts and it is the first  short gamma-ray burst to have 
been synchronously observed. 

This faint burst has been recorded FERMI observatory at 19:13:46.1 UT on 20
March 2009 at $ \alpha=10^h 22^m.8$, $ \delta=33d 24 ' $ with
with an uncertainty of 26.82 degrees (including the 3- degree systematic
error) \cite {gcn9056}. It is the faintest burst so far recorded by our cameras and
had therefore its coordinates determined with large errors. We found its upper limit to be
$V<9.1^m $ on 1 sec. images and
$ \frac {F_{opt}^{grb090328B}}{F_\gamma^{grb090328B}} < \frac {1}{1527} $ \cite
{gcn9065}

\begin{table}[!t]

\begin{longtable}{ p{2.2cm} p{2.3 cm} p{2.3 cm} p{1.5cm} p{3.0cm} p{1.8cm}
p{1.8cm} }

\caption {Prompt grb observations} \endhead

\hline \hline
GRB & Satellite & Site & V  & Optical fluence & Gamma-ray fluence (15-150keV) & Gamma-ray
fluence (8-1000keV) \\
\hline
GRB990123 & BeppoSAX  Konus-Wind & ROTSE-I  & 8.9$^m$ & $4.9 \cdot 10^{-7}$ &
$2.1 \cdot 10^{-5}$ &$3.0 \cdot 10^{-4}$ \\
GRB080319B & Swift & Pi-of-the-sky  TORTORA & 5.3$^m$ & $3.7 \cdot 10^{-6}$ &
$8.3 \cdot 10^{-5}$ &$4.4 \cdot 10^{-4}$ \\
\hline
GRB081102 & Swift & MASTER Kislovodsk & $<$13.0$^m$ & $<$ $2.8 \cdot 10^{-8}$ &
$2.3 \cdot 10^{-6}$ &$3.6 \cdot 10^{-6}$ \\
GRB090424 & Swift & MASTER Irkursk & $<$ 9.7$^m$ & $<$ $4.3 \cdot 10^{-8}$ &
$2.3 \cdot 10^{-5}$ &$5.2 \cdot 10^{-5}$ \\
GRB0901130B & FERMI & MASTER Kislovodsk & $<$ 11.0$^m$ & $<$ $4.1 \cdot 10^{-9}$
& $1.0 \cdot 10^{-6}$ &$2.0 \cdot 10^{-6}$ \\
GRB090305B (80\% part) & FERMI & MASTER Kislovodsk & $<$ 9.5$^m$ & $<$ $2.3
\cdot 10^{-9} $ & $2.7 \cdot 10^{-7}$ &$2.7 \cdot 10^{-6}$ \\
GRB090320B (80\% part) & FERMI & MASTER Kislovodsk & $<$ 11.0$^m$ & $<$ $1.5
\cdot 10^{-8} $ & $8.5 \cdot 10^{-7}$ &$1.1 \cdot 10^{-6}$ \\
GRB090328B (20\%  part) & FERMI & MASTER Irkursk & $<$ 9.1 $^m$ & $<$ $1.4 \cdot
10^{-10} $ & $2.1 \cdot 10^{-7}$ &$9.6 \cdot 10^{-7}$ \\

\hline\hline
\end{longtable}
\label {tab_grbwf}
\end{table}

\subsubsection{ Summary }

Figure~\ref{allgrb} and table~\ref{tab_grbwf} summarise the results of
observations of prompt emission of gamma-ray bursts. Even a cursory analysis 
of Fig. \ref{allgrb} demonstrates that optical emission is better correlated with soft 
rather than hard gamma-ray emission. The gamma-ray burst GRB090424 is located in
a remarkable position in Fig.~\ref {allgrb}. One can see that the upper limit for 
its optical emission is about one order of magnitude lower than the flux from the 
famous gamma-ray burst GRB990123, a pattern that may be indicative of irregular conversion
of gamma-ray emission into optical radiation from one burst to another. On the whole, it is safe to
conclude that more observations are required for a detailed analysis. The example
of the GRB090424 gamma-ray burst indicates that even the measured optical limit may constitute 
a result of real importance and that and this burst deserves separate analysis.

In this article we describe the methods of observation and analysis of images taken
with very wide-field cameras, and their application to synchronous observations of 
prompt gamma-ray burst emission in the MASTER-VWF experiment. We finally recall
that the renowned astrophysicist Bohdan Paczynski \cite{pach} pointed out in his last paper
the exceptional astrophysical significance of robotic telescope networks, which, like MASTER,
are composed of wide field searching cameras and larger telescopes. Now his and our ideas
at last become reality. 

This work was supported by the Ministry of Science of the Russian Federation (state contract N 02.740.11.0249 ).


\newpage

\begin{thebibliography}{}



\bibitem {master} Lipunov, V. M., Kornilov, V. G., et al. "Optical observations
of gamma-ray bursts, the discovery of supernovae 2005bv, 2005ee, and 2006ak, and
searches for transients using the "MASTER" robotic telescope" Astronomy Reports,
Volume 51, Issue 12, pp.1004-1025, 2007
\bibitem{lipunovmaster} Vladimir Lipunov, Victor Kornilov, Evgeny Gorbovskoy  et
al.  "MASTER ROBOTIC NET", eprint arXiv:0907.0827, Advances in astronomy in
press, 2009
\bibitem{mastergrbs} Tyurina Nataly,  Vladimir Lipunov, Kornilov Victor  et al. 
"MASTER prompt and follow-up GRB observations" eprint arXiv:0907.1036,  Advances
in astronomy in press, 2009
\bibitem {pach} Paczynski, Bohdan, "Astronomy with Small Telescopes" The
Publications of the Astronomical Society of the Pacific, Volume 118, Issue 850,
pp. 1621-1625., 2006
\bibitem {racusin} Racusin, J. L.; Karpov, S. V. et al. "Broadband observations
of the naked-eye γ-ray burst GRB080319B" Nature, Volume 455, Issue 7210, pp.
183-188, 2008
\bibitem {favor} S. Karpov, G. Beskin, A. Biryukov et al. "Optical camera with
high temporal resolution to search for transients in the wide field", Nuovo
Cimento C, issue 04-05, pp. 747-750, 2005
\bibitem {tortora} Molinari, E., Bondar, S., Karpov et al. "TORTOREM:
two-telescope complex for detection and investigation of optical transients",
Nuovo Cimento B, vol. 121, issue 12, pp. 1525-1526, 2006.
\bibitem {panat} A. Panaitescu, "Prompt GeV emission in the synchrotron
self-Compton model for Gamma-Ray Bursts" arXiv:0811.1235, 2008
\bibitem{sext} 	Bertin, E. and Arnouts, “SExtractor: Software for source
extraction.”,	Astronomy and Astrophysics Supplement, v.117, p.393-404, 1996.
\bibitem {widget} Onda, K.; Tamagawa, T.; Tashiro, M. et al. "Ultra wide-field
telescope WIDGET for observing GRB"  Il Nuovo Cimento B, vol. 121, Issue 12,
p.1549-1550, 2006
\bibitem {pi_of} M. Cwiok, W. Dominik, K. Malek et al., "Search for GRB related
prompt optical emission and other fast varying objects with "Pi of the Sky"
detector"Astrophysics and Space Science, Volume 309, Issue 1-4, pp. 531-535,
2007
\bibitem {zaspos} Zasov, Postnov "General astrophysics", Moscow 2006 in Russia
only.

\bibitem {gcn8516} E. Gorbovskoy, V. Lipunov, V.Kornilov, et al., "GRB 081102:
MASTER refind and final results" GCN Circular 8516, 2009 
\bibitem {gcn8471}  V. Lipunov, V.Kornilov,E. Gorbovskoy, et al., "GRB 081102:
MASTER prompt optical limit" GCN Circular 8471, 2008
\bibitem {gcn8597} E. Gorbovskoy, V. Lipunov, V.Kornilov, et al., "GRB 081130B:
MASTER prompt optical observations" GCN Circular 8597, 2008
\bibitem {gcn9252} E. Gorbovskoy, V. Lipunov, V.Kornilov, et al., "GRB 090424:
MASTER-Net prompt optical limit" GCN Circular 9252, 2009
\bibitem {gcn9233} E. Gorbovskoy, V. Lipunov, V.Kornilov, et al., "GRB 090424:
MASTER-Net prompt optical observations " GCN Circular 9233, 2009
\bibitem {gcn9004} E. Gorbovskoy, V. Lipunov, V.Kornilov, et al., "GRB090305B:
MASTER-net prompt optical short burst observations " GCN Circular 9004, 2009
\bibitem {gcn9038} E. Gorbovskoy, V. Lipunov, V.Kornilov, et al., "GRB 090320B:
MASTER-Net prompt optical observations" GCN Circular 9038, 2009
\bibitem {gcn8585} E. Gorbovskoy, V. Lipunov, V.Kornilov, et al., "GRB 081130:
MASTER VWF prompt optical observations Fermi GRB" GCN Circular 8585", 2009
\bibitem {gcn9065} K.Ivanov, S.Yazev, E. Gorbovskoy et al., "GRB 090328B:
MASTER-Irkutsk prompt optical short burst observations " GCN Circular 9065",
2009

\bibitem {gcn8468} E. E. Fenimore, S. D. Barthelmy, W. H. Baumgartner, et al.,
"GRB 081102: Swift-BAT refined analysis" GCN Circular 8468, 2008
\bibitem {gcn9231} T. Sakamoto, S. D. Barthelmy, W. H. Baumgartner, et al., "GRB
090424: Swift-BAT refined analysis" GCN Circular 9231, 2009
\bibitem {gcn9234} P.Schady and J. K. Cannizzo, "Swift/UVOT observations of GRB
090424" GCN Circular 9234, 2009

\bibitem {gcn8972} Colleen A. Wilson et al. "GRB090305B: Fermi GBM detection"
GCN Circular 8972, 2009
\bibitem {gcn9020} P. N. Bhat et al. " GRB 090320B: Fermi GBM detection" GCN
Circular 9020, 2009
\bibitem {gcn9230}  Valerie Connaughton et al. "GRB 090424: Fermi GBM
Observation" GCN Circular 9230, 2009
\bibitem {gcn9056}  A. Goldstein et al. "GRB 090328B: Fermi GBM Detection" GCN
Circular 9056, 2009


8674


MASTER VWF prompt optical observations Fermi GRB" GCN Circular 8585, 2008
 \bibitem {gcn8470} V. Mangano, B. Sbarufatti, V. La Parola, et al "GRB 081102:
Swift-XRT refined analysis" GCN Circular 8470, 2008
 \bibitem {8593} A.J. van der Horst, et al "GRB 081130B: Fermi GBM detection"
GCN Circular 8593, 2008


 \bibitem {calb} Kalberla, P. M. W.; Burton, W. B.; Hartmann, Dap et al. "The
Leiden/Argentine/Bonn (LAB) Survey of Galactic HI. Final data release of the
combined LDS and IAR surveys with improved stray-radiation corrections",
Astronomy and Astrophysics, Volume 440, Issue 2, pp.775-782, 2005

\end{thebibliography}
\end{document}